\documentclass[11pt]{article}
\pdfoutput=1
\usepackage{amsmath,amssymb}
\usepackage{graphicx,color}
\usepackage{braket}
\usepackage{mathtools}

\numberwithin{equation}{section}

\begin{document}

\begin{titlepage}
\thispagestyle{empty}

%\begin{flushright}
%XXXX \\
%YYYY \\
%\end{flushright}

\vspace{.4cm}
\begin{center}
\noindent{\Large \textbf{Strong subadditivity and holography}}\\
\vspace{2cm}

Andrea Prudenziati
%autor 2 etc... $^{a,b}$

\vspace{1cm}
  {\it
Institute of Physics, University of $S\tilde{a}o$ Paulo \\
05314-970 $S\tilde{a}o$ Paulo, Brazil\\
\vspace{0.2cm}
% $^{b}$institute 2\\
%address 2 etc...\\
\vskip 1ex
{\tt prude@if.usp.br }
 }

\vskip 2em
\end{center}

\vspace{.5cm}
\begin{abstract}
We study in detail the relationship between strong subadditivity for a boundary field theory and energy conditions for its bulk dual in 2+1 dimensions. We provide a discussion of known facts and new results organized from the simplest case of a static system with collinear intervals to a time dependent one in a generic configuration, with particular focus on the holographic geometric description. 
\end{abstract}

\end{titlepage}

\newpage

%\begin{scriptsize}
%\tableofcontents
%\end{scriptsize}

\newpage

\tableofcontents 

\section{Introduction}

Given an initial quantum state $\rho$, that may be either a pure state or a density matrix, and assuming factorizability of the total Hilbert space to which it belongs, $\mathcal{H}=\mathcal{H}_{A}\otimes\mathcal{H}_{B}\otimes\mathcal{H}_{C}\otimes\dots$, we first define the reduced density matrix $\rho_{A}$ as $\rho_{A}= Tr_{\mathcal{H}_{A}^{c}}\rho$, where $\mathcal{H}_{A}^{c}=\mathcal{H}_{B}\otimes\mathcal{H}_{C}\otimes\dots$ is the complementary of $\mathcal{H}_{A}$ in $\mathcal{H}$. The Entanglement Entropy ( EE ) is the Von Neumann entropy of $\rho_{A}$:
\[
S_A=-Tr_{\mathcal{H}_{A}}(\rho_A\log\rho_A).
\]
 It is easy to show that $S_A$ corresponds to some measure of the entanglement between degrees of freedom inside  $\mathcal{H}_{A}$ and degrees of freedom outside, hence the name. 

A simple formula exists for the holographic computation of EE, if factorizability of the total Hilbert space is induced by dividing the space ( or more covariantly a space-like surface ) into non intersecting subspaces $A$, $B$, $C$, $\dots$, each one defining the corresponding Hilbert space $\mathcal{H}_{A}$, $\mathcal{H}_{B}$, $\mathcal{H}_{C}$, $\dots$ for the corresponding local degrees of freedom, and if an holographic description of the quantum mechanical theory is possible in terms of Einstein gravity, with $\rho$ represented by a classical metric solution. Ryu and Takayanagi for the static case at fixed boundary time \cite{Ryu:2006bv}, and later Hubeny, Rangamani and Takayanagi ( HRT ) for the covariant generalization \cite{Hubeny:2007xt}, proposed the following:
\begin{equation}\label{ryutak}
S_A=\frac{\mathcal{A}(\Sigma(A))}{4 G_N}.
\end{equation}
In the above equation $G_N$ is the Newton constant in the gravitational $d+1$ dimensional theory, and $\mathcal{A}(\Sigma(A))$ is the area of the extremal codimension two space-like surface $\Sigma(A)$ homologous to $A$ and such that, at the boundary of the gravitational manifold, $\partial \Sigma(A)=\partial A$. If the gravitational geometry is static, the above formula simplifies in the Euclidean signature by using a minimal surface \footnote{We will call the formula (\ref{ryutak}) and the corresponding surface $\Sigma(A)$ as Ryu-Takayanagi or HRT depending if used for respectively static spacetimes in Euclidean signature or generically Lorentzian time dependent.}. In the static case a proof for the Ryu-Takayanagi formula was provided by \cite{Lewkowycz:2013nqa} \footnote{Although based on some assumptions, notably on the analytic continuation for the geometry dual to the replica trick, see \cite{Prudenziati:2014tta}.}, but still lacks for the time dependent covariant case. Nonetheless we will assume, throughout the paper, that (\ref{ryutak}) is valid.

One interesting thing about holography, and in particular when dealing with formulas like (\ref{ryutak}), is the interplay between quantum mechanics and classical general relativity; working with EE we can compare properties obeyed by $S_A$, deeply related to entanglement at the very foundations of quantum mechanics, with measures of areas of surfaces probing classical geometries of the dual space-time! The literature somehow related to this is vast, and covers the question of what quantum conditions ( inequalities ) can be derived from holography  \cite{Bao:2015boa}, \cite{Bao:2015bfa} and \cite{Hayden:2011ag}, the opposite problem of what gravitational restrictions are imposed from quantum inequalities \cite{Lashkari:2014kda} and \cite{Lin:2014hva}, studies the appearance of gravity equations of motions and dynamics from EE \cite{Banerjee:2014oaa}, \cite{Banerjee:2014ozp}, \cite{Faulkner:2013ica},\cite{Lashkari:2013koa} and \cite{Swingle:2014uza}, and the reverse \cite{Bhattacharya:2013bna} and \cite{Nozaki:2013vta}, and even reconstructs the metric from boundary EE data \cite{Spillane:2013mca}.

 It is well known that EE satisfies a series of inequalities, among which the two most restricting go under the name of strong subadditivity. Given three Hilbert spaces that factorize the total one $\mathcal{H}=\mathcal{H}_{A}\otimes\mathcal{H}_{B}\otimes\mathcal{H}_{C}\otimes\dots$, the two inequalities read:
\begin{subequations}
\begin{align}
S(A\cup B)+S(B\cup C) &\geq  S(A)+S(C) \label{ssa1}  \\
S(A\cup B)+S(B\cup C)&\geq  S(B)+S(A\cup B\cup C) \label{ssa2}  
\end{align}
\end{subequations}
where $S(A\cup B)$ refers to the EE computed by using the reduced density matrix living on the product of the two Hilbert spaces $\mathcal{H}_{A}\otimes\mathcal{H}_{B}$, and so on. If factorizability can be achieved by considering spacial non intersecting regions $A$, $B$ and $C$, then the same result holds for the corresponding EEs and $S(A\cup B)$ is the EE corresponding to the spatial region $A\cup B$ and so on. Note that these two equations are equivalent, as will be shown in the next section. Even if quantum mechanical proofs of (\ref{ssa1}) and (\ref{ssa2}) exist, see for example \cite{Chuang}, it is a nontrivial question to ask if the EE computed holographically by the Ryu-Takayanagi ( or covariantly HRT ) formula satisfies them or not. Why the question is nontrivial is due to three possible pitfalls. The first problem is the non induced factorizability of the total Hilbert space from dividing the physical space into subregions. It is well known for example for gauge theories, where gauge invariance at the boundary between a region $A$ and its complementary $A^c$ does not allow a factorization of the total Hilbert space of physical gauge invariant states into $\mathcal{H}_{A}$ and $\mathcal{H}_{A^c}$. Or, said in another way, the degrees of freedom of gauge theories are not point-like but rather nonlocal ( Wilson loops ), and consequently when considering a certain region of space they do not construct a physical ( gauge invariant ) Hilbert space. The literature is wide, see for example the lists of papers \cite{Aoki:2015bsa} for a more theoretical approach and \cite{Buividovich:2008gq} for attempts to formulate a lattice definition. 
While a formulation of EE may still be possible for gauge theories, for example by appropriately enlarging the Hilbert space or by correctly defining the path integral replica trick, whenever the total Hilbert space does not factorize we are in general not guaranteed that inequalities like (\ref{ssa1}) and (\ref{ssa2}) hold \footnote{ As far as I know it is believed that gauge theories  satisfy strong subadditivity, but the discussion here is general. }. In fact this argument may apply even to EE computed non holographically, as long as the reduced density matrices are defined with respect to a certain region of space.

 A pure CFT example of strong subadditivity violation without passing through the holographic description, but rather considering EE computed using the replica trick in the CFT, will be provided in section \ref{cft}. There we will consider two dimensional CFTs with Lorentz anomaly, by using the results of \cite{Castro:2014tta} where it was noted that the EE, as expected, is not Lorentz invariant. This means that the EE transforms under boosts and, if these are applied to space regions entering inequalities (\ref{ssa1}) and (\ref{ssa2}), the two sides can change so that to possibly lead to a violation. Said otherwise, strong subadditivity may be respected in some fixed reference frame in which the quantization of the theory has been implemented, but if we consider Hilbert spaces corresponding to space regions that have been differently boosted with respect to this original frame, and the theory has a Lorentz anomaly, violation can occur. This is the second problem and again it applies even without passing through an holographic computation. On the issue of EE and its relation to anomalies see also the recent papers \cite{Nishioka:2015uka}  \cite{Iqbal:2015vka} \cite{Hughes:2015ora}.

The third problem arises only in holographic EE: while Einstein's equations can be solved using any energy momentum tensor as a source, it is not guaranteed that any choice is actually physically meaningful. To constrain the energy momentum tensor various energy conditions have been proposed, see for example \cite{Hawking:1973uf} for a review of these conditions and of their effects. Without imposing any condition it is not improbable that the resulting classical geometry may not be dual to same actual physical quantum system. In particular we are no longer guaranteed on the validity of any quantum mechanically-derived inequality \footnote{ We are here necessarily vague on the meaning of physical. }. We will discuss this further in the next section in relationship with the proof of the equivalence between  (\ref{ssa1}) and (\ref{ssa2}), and in the conclusions. 

Note that, if any of the above pitfalls affects the proof of either (\ref{ssa1}) or (\ref{ssa2}), that is we have violation of only one of the two inequalities, then similar reasons should necessary lead to the violation of the proof for the equivalence between (\ref{ssa1}) and (\ref{ssa2}) that we will discuss in the next section.

Extremely simple holographic proofs of (\ref{ssa1}) and (\ref{ssa2}) exist when the three connected adjacent regions are at the same constant boundary time  ( or its boosted version ) and the bulk geometry is static, somehow in contrast with the complication of the standard quantum mechanical proofs. However the argument fails when more generic configurations of regions and/or time dependent backgrounds are considered. Because of this some works considered dropping one or both of these restrictions in order to check if the strong subadditivity inequalities were still satisfied or not, and if some conditions should be eventually applied to the geometry in order to enforce their validity. In particular \cite{Lashkari:2014kda} and \cite{Lin:2014hva} considered static backgrounds with generic space-like boundary regions $A$, $B$ and $C$, and found that requiring (\ref{ssa2}) leads to a integrated version of the Null Curvature Condition ( NCC ) (\ref{intncc}). Further a simple time dependent geometry, asymptotically AdS in 2+1 dimensions which is the Vaidya metric was used by \cite{Allais:2011ys}, \cite{Callan:2012ip} and  \cite{Caceres:2013dma}; depending on the choice of a sign the Vaidya geometry may be selected to satisfy or not the local NCC (\ref{ncc}), and what was found is that the NCC is a sufficient requirement to respect (\ref{ssa2}), while (\ref{ssa1}) is always satisfied. Finally a proof for (\ref{ssa2}) was provided by \cite{Wall:2012uf} for dual geometries satisfying the NCC  and generic connected adjacent space-like intervals. 

The main goals of this paper are three. First of all to review in an organized way the present understanding of strong subadditivity inequalities in holographic theories and their connection with energy conditions. Second to enlighten the geometrical part of the above problem studying what is the holographic counterpart of the violation of strong subadditivity and the role of energy conditions. Third to fill some gaps in the literature and discuss further discoveries as, notably, the violation of (\ref{ssa2}) by two dimensional CFTs with Lorentz anomaly, the development of time-like distances between geodesics entering  (\ref{ssa2}) whenever the inequality does not hold and some new proofs along the way. 

The paper is organized by discussing strong subadditivity in set-ups of crescent complication, always in 2+1 dimensions in the bulk ( although some results may be generalized ) and boundary regions $A$, $B$ and $C$ chosen to be adjacent, starting with static backgrounds with collinear intervals then moving to generic space-like configurations and finally to time dependent geometries, discussing both the purely quantum mechanical problem and its holographic version.  Two appendices contain respectively the generic result and proof of what configurations for the boundary intervals  $A$, $B$ and $C$ create the strongest bound on EE by strong subadditivity, and various computations for Vaidya metric that will be used in the last sections as specific examples for the more generic discussion. 

\section{A few facts on strong subadditivity}\label{init}

We begin with the purely quantum mechanical proof of the equivalence between the two strong subadditivity inequalities, (\ref{ssa1}) and (\ref{ssa2}) \cite{Chuang}. Let us introduce an auxiliary Hilbert space $\mathcal{H}_D$ such that partial tracing some pure state $\ket{\Psi}$ over it, reproduces the reduced density matrix $\rho_{A\cup B\cup C}$ \cite{Chuang}:
\begin{equation}\label{12equiv}
\rho_{A\cup B\cup C}= Tr_{\mathcal{H}_D}\ket{\Psi}\bra{\Psi}.
\end{equation}
Then
\[
S(A\cup B\cup C)=S(D) \;\;\; S(B\cup C)=S(A\cup D)
\]
and we can convert (\ref{ssa1}) into (\ref{ssa2})
\[
S(A\cup B)+S(B\cup C)=S(A\cup B)+S(A\cup D)\geq S(B)+S(D)=S(B)+S(A\cup B\cup C).
\]
A discussion on why this proof may fail in certain circumstances can be found inside \cite{Allais:2011ys}. The argument is that, purely quantum mechanically we are always guaranteed to find $\mathcal{H}_D$.  If the theory has zero entropy it is just the complementary of $\mathcal{H}_{A}\otimes\mathcal{H}_{B}\otimes\mathcal{H}_{C}$ inside the total Hilbert space, $\mathcal{H}_{D}=(\mathcal{H}_{A}\otimes\mathcal{H}_{B}\otimes\mathcal{H}_{C})^c$, while if the theory is described by some density matrix it may not be the Hilbert space of degrees of freedom of the theory itself. Still the proof applies if we limit ourselves to EE computed quantum mechanically and not holographically. The only pitfall in this case is the already discussed issue of factorizability of the total Hilbert space that we will not repeat.  Holographically instead, if the bulk theory contains a black hole, the total entropy of the system in non zero and the boundary theory is not in a pure state but rather in a thermal density matrix. The proof then holds for the holographic entanglement entropy only if the bulk actually describes some quantum system and we can thus use the above derivation. If the bulk is sufficiently unphysical we may instead have troubles; this is in fact just another way of recasting the argument that generic bulk geometries may not be dual to quantum mechanical systems, as discussed in the introduction. Finally we will soon see a purely CFT violation of (\ref{ssa2}) in the following sections while preserving (\ref{ssa1}) for two dimensional theories with Lorentz anomaly, so the above proof should break down in this case for analogous reasons. The argument here is as in the introduction, and the problem arises when we consider differently boosted space regions starting from the reference frame that was used to quantize the theory and in which strong subadditivity holds; Lorentz anomaly implies that all the expressions written in the original frame will transform and relationships between them are no longer guaranteed to hold.

Similarly simple proofs do not exist for the two strong subadditivity inequalities, that from now on we will name SSA1 for (\ref{ssa1}) and SSA2 for (\ref{ssa2}), that instead require some amount of work to be verified. Simplicity is restored when we consider their holographic description as Ryu-Takayanagi surfaces when considering collinear boundary intervals in static spacetimes.

\begin{figure}[h]
\centering
\vspace{-0pt}
\includegraphics[width=0.7\textwidth]{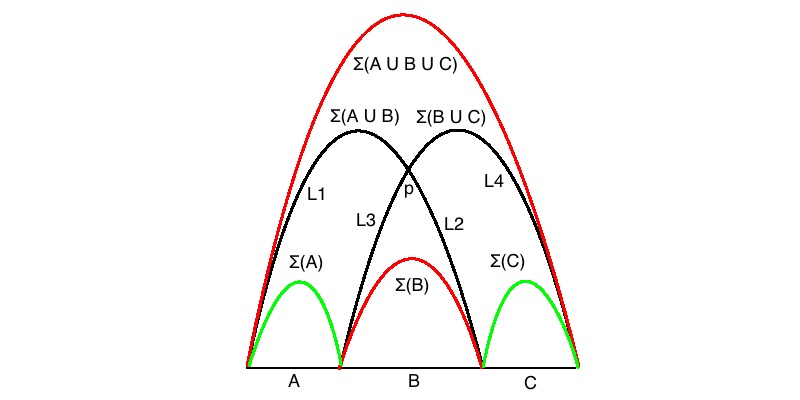}
\vspace{-0pt}
\caption{Static proof for collinear intervals.}
\label{figuraspc}
\end{figure}

For the rest of the paper, unless otherwise stated, the 2+1 dimensional bulk theories will be in the Euclidean signature when static, and Lorentzian when dynamic. The easiest case is that of static geometry with the three boundary intervals collinear, either at fixed time or belonging to a straight space-like line; also we assume Lorentz invariance. It is quite surprising that the proof for both (\ref{ssa1}) and (\ref{ssa2}) just amounts to look at picture \ref{figuraspc}. The one dimensional minimal surfaces $\Sigma(A\cup B)$ and $\Sigma(B\cup C)$ intersect at the point p, thus defining $L_1$, $L_2$, $L_3$ and $L_4$; due to the minimality of $\Sigma(A)$ and $\Sigma(C)$, it is immediate the result $\mathcal{A}(\Sigma(A)) \leq \mathcal{A}(L_1 \cup L_3)$ and $\mathcal{A}(\Sigma(C)) \leq \mathcal{A}(L_2 \cup L_4)$, proving (\ref{ssa1}). Analogously we can obtain (\ref{ssa2}) from $\mathcal{A}(\Sigma(A\cup B\cup C)) \leq \mathcal{A}(L_1 \cup L_4)$ and $\mathcal{A}(\Sigma(B)) \leq \mathcal{A}(L_2 \cup L_3)$.
The above proofs unfortunately only apply to the case described. If $A$, $B$ and $C$ for example are not collinear, and/or if the geometry is time dependent so that the curves bend in the time direction, then we will not generically have any intersection between $\Sigma(A\cup B)$ and $\Sigma(B\cup C)$. Further, the time dependent case brings one additional complication as the HRT surfaces are in this case extremals, so even if we could actually find a way to compare areas, we would not be able to write down inequalities as if minimal surfaces were involved.

\section{Static case}

\subsection{Monotonicity and concavity}\label{mc}

When Lorentz invariance is preserved the EE is just a function of the proper length of the interval $S=S(l)$;  SSA1 implies monotonicity for the function $S=S(l)$ while for collinear intervals SSA2 leads instead to concavity. Monotonicity is immediately proven by considering the special case of collinear intervals with proper lengths $l(A)=l(C)=l$ and $l(B)=d$, then (\ref{ssa1}) just gives 
\begin{equation}\label{mon}
S(l+d)\geq  S(l).
\end{equation}
In fact it is also obviously true the opposite, that monotonicity implies SSA1 and the two conditions are equivalent. When considering EE computed holographically by the area of Ryu-Takayanagi surfaces, monotonicity can in fact be obtained directly without passing through SSA1, by showing that the proper length of a minimal surface ( or more generally any curve that minimize some fixed bulk functional ) is monotonically increasing as a function of the proper length of the boundary interval to which it is attached. 

\begin{figure}[h]
\centering
\vspace{-0pt}
\includegraphics[width=0.7\textwidth]{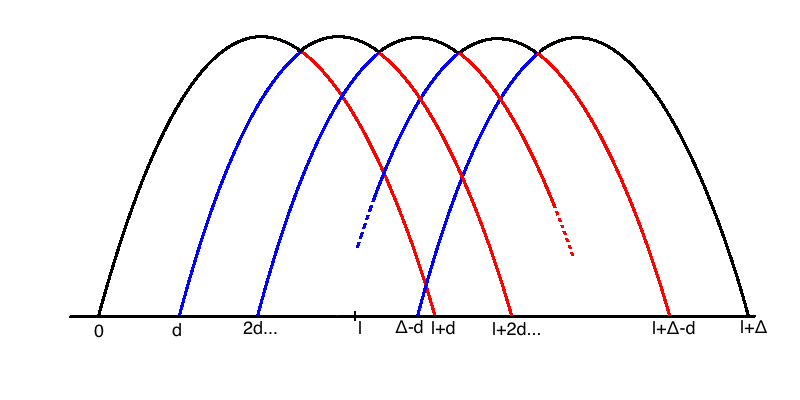}
\vspace{-20pt}
\caption{Proof of concavity for minimal surfaces.}
\label{figuramonmin}
\end{figure}

Concavity is just barely more complicated to derive from SSA2, see for example \cite{Callan:2012ip} for the proof of the equivalence between concavity and SSA2 for collinear intervals. Again this property can be obtained directly, as for monotonicity, when dealing with the holographic minimal surfaces $\Sigma(l)$, whenever we vary the proper length $l$ of the boundary interval along the fixed space-like direction determined by the boundary end points. This is obtained with a slightly different construction then the one of figure \ref{figuraspc}; let us consider a boundary straight space-like line and pick $\Delta/d=n$ ( $n \in \mathbb{Z} \gg 1$ ) minimal curves ending on intervals of length $l+d$ that belongs to such line, displaced by a distance $d$, as in figure \ref{figuramonmin} ( $l$ does not need to be an integer multiple of $d$ ). 
We have that the proper length of $\Sigma$, $\mathcal{A}(\Sigma)$, obeys the inequality:
\[
\frac{\Delta}{d}\;\mathcal{A}(\Sigma(l+d))\geq \left( \frac{\Delta}{d}-1 \right)\mathcal{A}(\Sigma(l))+\mathcal{A}(\Sigma(l+\Delta))
\]
which is proven as in figure \ref{figuramonmin}, just noticing that the black curve has necessarily higher or equal proper length ( or whatever functional we are minimizing ) than $\Sigma(l+\Delta)$, while  each of the $\Delta/d-1$ unions of a blue and a red arc is bigger than the proper length of what would be the corresponding minimal curve $\Sigma(l)$. In the limit of $d \rightarrow 0$, while keeping $\Delta\geq 0$ at some finite value, the above formula reads
\begin{equation}
\frac{\mathcal{A}(\Sigma(l+d))-\mathcal{A}(\Sigma(l))}{d}\Delta+\mathcal{A}(\Sigma(l))\geq \mathcal{A}(\Sigma(l+\Delta)) \xRightarrow{d\rightarrow 0} \mathcal{A}(\Sigma(l))+\Delta \;\partial_m\mathcal{A}(\Sigma(m))\Big|_{l}\geq \mathcal{A}(\Sigma(l+\Delta)) 
\end{equation}
that is just concavity for $\mathcal{A}(\Sigma(l))$.

The case of non collinear intervals is more interesting. In Appendix \ref{A}  it is shown that, as a function of the slopes of the three intervals $\alpha_A,\alpha_B,\alpha_C$, the stricter bound on the EE $S(l)$ from SSA1 comes from the case $\alpha_A\geq0,\alpha_B=1,\alpha_C\geq 0$, and from SSA2 $\alpha_A=1,\alpha_B\geq 0,\alpha_C=-1 $ or $\alpha_A=-1,\alpha_B\geq 0,\alpha_C=1 $ ( or their parity transformed counterparts ). 

Let us show what conditions on $S(l)$ the strong subadditivity inequalities corresponds to for these configurations. The SSA2 inequality has already been considered in \cite{Casini:2004bw} for a configuration slightly less general then the one we are using, and we will follow a similar procedure. For reference for the parameterization consider the first picture of figure \ref{figuravssa} ( the situation with $\alpha_A=-1,\alpha_B\geq 0,\alpha_C=1 $ is completely analogous ); defining $a\equiv \log r$, $b\equiv \log s$ and $a_x\equiv \log(r-2x)$, $b_y\equiv \log(s-2y)$, and a function $G$ such that $G(a)\equiv S(e^\frac{a}{2})$, by computing the proper lengths for the intervals appearing in SSA2, (\ref{ssa2}) reads for this case
\[
G(a+a_x)+G(b+b_y)\leq G(b+a_x)+G(a+b_y).
\]
As we always have $b_y> a_x$ ( as explained in figure \ref{figuravssa} caption ) and obviously $b>a$, the above inequality is just concavity for $G(a)$, so
\begin{equation}\label{con2}
0\geq \partial^2_a G(a)=\partial_a(\partial_a S(e^\frac{a}{2})) \Rightarrow \partial^2_l S(l)\leq - \frac{\partial_l S(l)}{l}
\end{equation}
that is stronger than simple concavity for $S(l)$, due to the monotonicity required from SSA1: $\partial_l S(l)\geq 0$. 

We can now follow a similar path for SSA1 applied to the configuration with slopes $\alpha_A\geq0,\alpha_B=1,\alpha_C\geq 0$. In addition, as the interval B appears only on the bigger side of the inequality, an even stronger bound can be obtained by further minimizing the proper lengths of $A\cup B$ and $B\cup C$ by considering an infinitesimal space coordinate length $x_B\ll 1$. Parameterizing this configuration as in appendix \ref{A}, and defining $x_A\sqrt{1-\alpha_A^2}\equiv e^a$, $\frac{x_B}{x_A(1+\alpha_A)}\equiv \epsilon_a$ ( and correspondingly for $a\leftrightarrow c$ ) and $F(a)\equiv S(e^a)$, we obtain the following result from (\ref{ssa1}) ( using $x_B\ll 1$ to rewrite $\sqrt{1+2\epsilon_a} \approx e^{\epsilon_a}$ )
\[
F(a)+F(c)\leq F(a+\epsilon_a)+F(c+\epsilon_c)
\]
which is just monotonicity for $F(a)$. However in this case, as a first derivative is involved, the condition implies nothing more than the usual monotonicity for $S(l)$:
\begin{equation}\label{con1}
0\leq\partial_a F(a)=l\partial_l S(l) \Rightarrow \partial_l S(l)\geq 0.
\end{equation}
That (\ref{con2}) becomes stricter than concavity for non collinear intervals while (\ref{con1}) remains the usual monotonicity will soon have its counterpart: in various cases, holographically and not, for generic adjacent interval configurations EE will still satisfy SSA1 while the stricter SSA2 will be generically violated. In the holographic description we will see as respecting SSA2 shall require certain geometrical conditions to be satisfied by the background.

\subsection{First appearance of energy conditions in the bulk}\label{faec}

\begin{figure}[h]
\centering
\vspace{-20pt}
\includegraphics[width=1.0\textwidth]{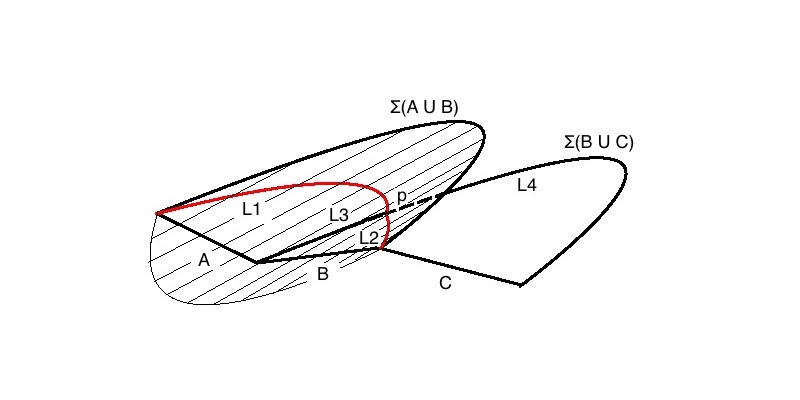}
\vspace{-50pt}
\caption{Partial picture for the geometric construction leading to the proof of the requirement of the integrated NCC for SSA2 for static non collinear intervals. Depicted is the null congruence $N(\Sigma(A\cup B))$ and its intersection $L_1\cup L_2$ with the ( not shown ) achronal slice.}
\label{figuraspnc}
\end{figure}

Still studying static bulk geometries, with all the nice properties listed in the previous sections but this time in the Lorentz signature, let us see what happens in the bulk when considering geodesics bounded to non collinear adjacent intervals, as for instance in figure \ref{figuraspnc}. The first clear issue is that geodesics ending on $A\cup B$ and $B\cup C$ do not generically intersect, and consequently try to prove SSA1 and SSA2 as in figure \ref{figuraspc} does not work. SSA1 however has an alternative proof derivable from the simple relation,  \cite{Lashkari:2014kda} and \cite{Myers:2012ed}:
\begin{equation}\label{r0}
\frac{dS(l)}{dl}=r_0
\end{equation}
where $r_0$ is the conformal scale factor of asymptotically AdS metrics, reached by the geodesics at its vertex ( coordinates chosen so that we have large $r$ when close to the boundary ).
As it will be useful later, let us review briefly the derivation of (\ref{r0}), following \cite{Lashkari:2014kda} with some modification. The assumptions are a conformal two dimensional boundary Minkowski metric and translation invariance along a boundary space-like  direction with associated Killing vector $\xi^{\mu}$. The geodesic extremizes the action
\[
S=\int_{i}^{f}d\lambda\sqrt{g_{\mu\nu}\partial_{\lambda}x^{\mu}\partial_{\lambda}x^{\nu}}
\]
where the interval $i-f$ is taken to be at fixed boundary time $t_b$. We vary only the position of the final end point "f " by a purely spatial translation $\delta x_f^{\mu}=\delta x\; \xi^{\mu} $. The variation of the action is just the boundary term $\delta S =p_{\mu}\delta x_f^{\mu}$ with $p_{\mu}\equiv \partial L/\partial (\partial_{\lambda}x^{\mu})$. Then
\[
\frac{\delta S}{ \delta x} =\xi^{\mu} p_{\mu}.
\]
As this quantity is conserved along all the geodesic we can evaluate it at its vertex, where $p_{\mu}=g_{\mu\nu}\partial_{\lambda}x^{\nu}/\sqrt{g_{\mu\nu}\partial_{\lambda}x^{\mu}\partial_{\lambda}x^{\nu}}$ simplifies by writing $\partial_{\lambda}x^{\mu}=\xi^{\mu} (\xi \cdot \partial_{\lambda}x)/(\xi^{\nu}\xi^{\rho}g_{\nu\rho})$. A brief computation leads to 
\begin{equation}\label{r0st}
\frac{\delta S}{ \delta x} =\sqrt{\xi^{\nu}\xi^{\rho}g_{\nu\rho}}=r_0.
\end{equation}
This equation is in fact true with or without time translation invariance. If time translation is a symmetry of the bulk, we can further boost the above equation (\ref{r0st}) to obtain (\ref{r0}).

The important point here is that $r_0>0$ always, and thus $S(l)$ is monotonically increasing. One may then try to use this formula to check (\ref{con2}) by exploiting the dependence  $r_0(l)$; the result obtained by \cite{Lashkari:2014kda} ( see also \cite{Lin:2014hva}  ) is that SSA2 in its stronger bound (\ref{con2}) is equivalent to the condition on the bulk geometry
\begin{equation}\label{intncc}
\int_{\Sigma} R_{\mu\nu} k^{\mu}k^{\nu}\geq 0
\end{equation}
where $\Sigma$ is our geodesic ending on the interval of proper length $l$ and $k^{\mu}$ is a null vector perpendicular to $\xi^{\mu}$ \footnote{This energy condition may also be written replacing $R_{\mu\nu}\leftrightarrow T_{\mu\nu}$, whenever Einstein's equations are holding.}. Our goal is to find a more geometrically transparent proof for the emergence of (\ref{intncc}), enhancing the difference between collinear and non collinear intervals and the role of $R_{\mu\nu} k^{\mu}k^{\nu}$, where the original derivation  uses a metric ansatz and Einstein's equation to show the equivalence between the explicit formulae for (\ref{intncc})  and (\ref{con2}).

So let us start  from figure \ref{figuraspnc}; the first issue to solve is that the two curves $\Sigma(A\cup B)$ and $\Sigma(B\cup C)$ do not generically intersect. Let us then shoot out from $\Sigma(A\cup B)$ ( that in our example has been chosen to be nowhere in the past of $\Sigma(B\cup C)$, if otherwise inverting the roles ) a congruence of null geodesics in the past boundary direction, that is then a codimension one surface $N(\Sigma(A\cup B))$. Given translation invariance in the boundary space direction, this null geodesics are chosen to be perpendicular to the corresponding Killing vector, this in order to ensure always intersection between $N(\Sigma(A\cup B))$ and $\Sigma(B\cup C)$ at a point $p$. Now consider any achronal slice ( with points either space-like or null separated ) that contains both the boundary interval $A\cup B$ and the point $p$, and intersects $N(\Sigma(A\cup B))$ along some curve; this curve goes from the left end point of $A$ to $p$, lets call this piece $L_1$, and continues from $p$ to the right end point of $B$, that we call $L_2$ as shown in figure \ref{figuraspnc}. The point $p$ also splits up $\Sigma(B\cup C)$ into $L_3$ to the left and $L_4$ to the right. Here enters the Raychaudhuri equation
\begin{equation}\label{ray}
\frac{d\Theta}{d\lambda}=-\Theta^2-\sigma_{\mu\nu}\sigma^{\mu\nu}-R_{\mu\nu} k^{\mu}k^{\nu}
\end{equation}
where, applied to our case, $\lambda$ is the affine parameter along $k^{\mu}$, $\sigma^{\mu\nu}$ is the shear and $\Theta$ represents the variation of the line element of $\Sigma(A\cup B)$ along $\lambda$, divided by the line element itself. As $\Sigma(A\cup B)$ is a geodesic, $\Theta=0$ on it; then, as on the right hand side of (\ref{ray}) all the quantities are negative definite but $R_{\mu\nu} k^{\mu}k^{\nu}$, the total proper length of $\Sigma(A\cup B)$ decreases along $\lambda$ if the integral of $R_{\mu\nu} k^{\mu}k^{\nu}$ on $\Sigma(A\cup B)$ is higher or equal then zero, which is just the integrated NCC condition  (\ref{intncc})  \footnote{The present argument would require the integrated NCC not only on $\Sigma$, but also on all the evolution curves created along the flow by $\lambda$. However we can require only the integral on $\Sigma$ if we restrict to boundary intervals with $A$ and $C$ of infinitesimal coordinate length. This is also what was done in \cite{Lashkari:2014kda}, although by a different road, and there it was also shown as having SSA2 respected for this infinitesimal configuration implies SSA2 valid for generic cases. So only integration along $\Sigma$ is actually required.}. We finally obtain
\begin{equation}\label{sss}
\mathcal{A}(\Sigma(A\cup B))+\mathcal{A}(\Sigma(B\cup C))\geq \mathcal{A}(L_1)+\mathcal{A}(L_2)+\mathcal{A}(L_3)+\mathcal{A}(L_4)\geq \mathcal{A}(\Sigma(B))+\mathcal{A}(\Sigma(A\cup B\cup C))
\end{equation}
where the first inequality is guaranteed by the Raychaudhuri equation coupled with (\ref{intncc}); the second inequality instead, is the usual argument of figure \ref{figuraspc}, but this time applied to extremal surfaces restricted to common achronal slices, one containing  $L_2$, $L_3$ and $\Sigma(B)$, and the other $L_1$, $L_4$ and $\Sigma(A\cup B\cup C)$. On such a slice extremal surfaces are minimal ( as will also happen when considering the maximin construction of \cite{Wall:2012uf} ), and that such slices exist on the present static case is what makes the proof applicable here but not on the corresponding time dependent situation, where the work will be harder. This proves that the integrated NCC implies (\ref {intncc}) for any adjacent interval configurations in static spacetimes. The reverse is true only for configurations maximizing the SSA2 bound, like the ones discussed in the previous section or in appendix \ref{A}. 

\subsection{$c_L\neq c_R$ theories and strong subadditivity violation}\label{cft}

It is now an interesting question to ask if we can find a purely CFT example of violation of strong subadditivity, using a time independent state $\rho$ ( the vacuum ), as to my knowledge so far all the cases found in the literature rely on the holographic description, often with time dependent metrics. By analysing its holographic counterpart we can isolate, inside the mechanics of strong subadditivity violation in the bulk, the dual of a genuine boundary CFT violation rather then issues of the holographic formula of EE for possibly unphysical backgrounds. 

In fact such an example exists and it is when a two dimensional CFT has different left and right central charges. The computation of EE for the vacuum was done in \cite{Castro:2014tta}, together with their proposal for the holographic counterpart that we will discuss in the next section, giving as a result for the EE in Lorentzian signature ( otherwise the result is complex )
\begin{equation}\label{clcree}
S(A)=\frac{c_L+c_R}{6}\log\left(\frac{l(A)}{\epsilon}\right)+\frac{c_R-c_L}{6}\;\alpha(A)
\end{equation}
where $\epsilon$ is the UV regulator and $\alpha(A)=$ ( or $\alpha_A$ as in the notation of appendix \ref{A} )  is the slope of the interval $A$ with respect to the constant time line, or equivalently the rapidity of the Lorentz boost from a constant time interval to the actual $A$.

\begin{figure}[h]
\centering
\vspace{-10pt}
\includegraphics[width=1.0\textwidth]{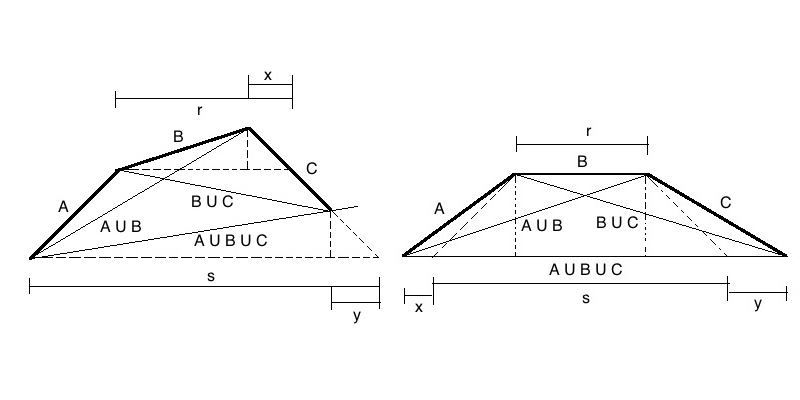}
\vspace{-40pt}
\caption{Two configurations of intervals for evaluation of SSA2. Oblique hatched lines are light-like, as well as A and C intervals in the first picture. Parameters $x,y$ are always positive in the second picture, while may be negative in the first with the obvious bounds for positive coordinate length for $C$, $(s-r)/2-y+x> 0$, and for space or light-like interval $B$, $x<r/2$. }
\label{figuravssa}
\end{figure}

Let us start with the inequality SSA2. We expect the highest amount of violation, if any, by some of the interval configurations that maximize the bound from SSA2, as explained in appendix \ref{A}. In fact, as the term proportional to $c_L+c_R$ inside (\ref{clcree}) respects SSA2 as it is the usual two dimensional EE result from the vacuum of ( non-Lorentz violating ) CFTs, to have SSA2 violation we have better to manage an interval configuration that saturates the inequality for that term, or is close enough ( note that the  UV regulator $\epsilon$ simplifies inside the strong subadditivity inequalities ). Among these configurations the simplest one has the interval $B$ at constant time $\alpha_B=0$, and $A$ and $C$ of the same proper length $l(A)=l(C)$ with light-like slopes $\alpha_A=1$, $\alpha_C=-1$ ( or $\alpha_A=-1$, $\alpha_C=1$ ). Using (\ref{clcree}) however, it is immediately clear that it respect SSA2 because $\alpha_{A\cup B}=-\alpha_{B\cup C}$ and $\alpha_B=\alpha_{A\cup B\cup C}=0$. We then consider some deformations of it, either by studying a generic bound-maximizing structure like the first one in figure \ref{figuravssa}, or one where the segment $A$ and $C$ are made space-like, as in the second picture of figure \ref{figuravssa} \footnote{ We could have just studied the most generic adjacent interval configuration and found violation, but the above examples, because of the parameterization used, make  the analysis more transparent. }.

Let us start from the first configuration; writing down the proper lengths of all the intervals as functions of the parameters $r,s,x$ and $y$ as appearing in the picture, SSA2 reads as follows
\[
\frac{c_L+c_R}{12}\left(\log (r(r-2x))+\log (s(s-2y))\right)+\frac{c_R-c_L}{6}\left(\alpha_B+\alpha_{A\cup B\cup C}\right)\leq \;\;\;\;\;\;\;\;\;\;\;\;\;\;\;\;\;\;\;\;\;\;\;\; 
\]
\[
\;\;\;\;\;\;\;\;\;\;\;\;\;\;\;\;\;\;\;\;\;\;\;\; \leq \frac{c_L+c_R}{12}\left(\log (s(r-2x))+\log (r(s-2y))\right)+\frac{c_R-c_L}{6}\left(\alpha_{A\cup B}+\alpha_{B\cup C}\right).
\]
The terms proportional to the sum of the central charges and containing the dependence of the EE on the interval's lengths simplify leaving an inequality that depends only on the angles $\alpha$s. As the tangent of an angle between $-\pi/2$ and $\pi/2$ is a monotonically increasing function, this inequality may be rewritten taking the tangent of the sums $\alpha_B+\alpha_{A\cup B\cup C}$ and $\alpha_{A\cup B}+\alpha_{B\cup C}$, and using the usual formula for $\tan{\alpha+\beta}$ we obtain
\[
\frac{c_R-c_L}{6}\frac{\tan{\alpha_B}+\tan{\alpha_{A\cup B\cup C}}}{1-\tan{\alpha_B}\tan{\alpha_{A\cup B\cup C}}}\leq \frac{c_R-c_L}{6}\frac{\tan{\alpha_{A\cup B}}+\tan{\alpha_{B\cup C}}}{1-\tan{\alpha_{A\cup B}}\tan{\alpha_{B\cup C}}}
\]
that in terms of $r,s,x,y$ reads
\[
c_L>c_R\;\;\;\;\;2 x y \leq s x + r y 
\]
\[
c_R>c_L\;\;\;\;\;2 x y \geq s x + r y.
\]
Every configuration out of this parameter range violates SSA2. 

\begin{figure}[h]
\centering
\vspace{+10pt}
\includegraphics[width=0.5\textwidth]{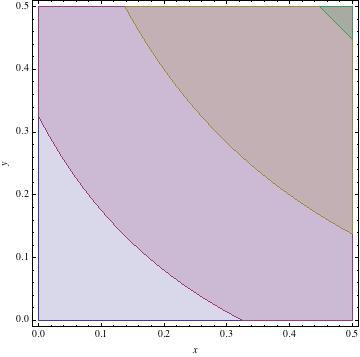}
\vspace{-0pt}
\caption{The coloured regions ( extending all the way to the upper right corner ) represent the $x,y$ parameter values respecting the inequality (\ref{2ssa2}) for $c_L=5,r=1,s=2$ and $c_R=5,4,3,2$ respectively; the region decreases as $c_R$ becomes smaller. For $c_R\geq c_L$  the inequality always holds in this example. }
\label{figurasecssa2}
\end{figure}

The second configuration instead leads to an SSA2 inequality that reads as follow:
\begin{equation}\label{2ssa2}
\frac{c_L + c_R}{6} (\log (s + x + y) + \log (r) ) \leq 
   \frac{c_L + c_R}{12}  (\log (r + y)(s + y) + \log(r + x)(s + x)) +
\end{equation}
\[
+\frac{c_R - c_L}{6}  \left(\arctan \left(\frac{s - r}{r + s + 2 y}\right) + 
   \arctan \left(\frac{s - r}{r + s + 2 x}\right)\right).
\]
The $x,y$ parameter region that respects the above inequality is shown in figure \ref{figurasecssa2} for a specific example. Violation is obvious for a region increasing as $c_R$ becomes smaller than $c_L$.

Finally we inspect SSA1 for a generic ( adjacent ) interval configuration parameterized as in figure \ref{figuracldcr} of appendix \ref{A}. The resulting inequality depends on the parameters $x_A,x_B,x_C$ and $\alpha_A,\alpha_B,\alpha_C$ and it contains both logarithms and arcotangents. We may split it in two, independent inequalities, proportional to $c_L$ and $c_R$ that can be analysed separately ( respecting both is a sufficient condition for respecting the complete equation ). We couldn't provide a proof, but numerical analysis for various values of $x_A,x_B,x_C$ and the complete range of  $\alpha_A,\alpha_B,\alpha_C$ shows that no violation of SSA1 occurs.

\subsection{$c_L\neq c_R$ discussion and holographic description}

In the introduction and section \ref{init} we have discussed the quantum mechanical mechanism that may lead to the strong subadditivity violation in Lorentz anomalous theories, as observed in the previous section. Also a detailed discussion on EE in anomalous theories can be found in \cite{Hughes:2015ora}, \cite{Iqbal:2015vka} and \cite{Nishioka:2015uka}. Here we want to understand how this mechanism acts when looking at the dual holographic theory.

The holographic description of CFTs with $c_L\neq c_R$ is provided by a theory called Topological Massive Gravity, TMG, with an action which is the sum of Einstein-Hilbert ( with possible cosmological constant ) and gravitational Chern-Simons with a real relative coefficient of $1/\mu$, see \cite{Castro:2014tta} and references within. The holographic description of EE for these theories has been proposed to be given by a curve $\Sigma$, with the usual boundary conditions and holonomy properties, but extremising a functional which is not its proper length but instead, in Lorentzian signature \footnote{In Euclidean signature an $i$ appears in front of the framing term and the functional becomes complex, as the corresponding boundary result (\ref{clcree}).},
\begin{equation}\label{nfun}
\int_{\Sigma}d\tau\left(\sqrt{g_{\mu\nu}\partial_{\tau}x^{\mu}\partial_{\tau}x^{\nu}}+\frac{1}{\mu}n_2^{\mu} (\bigtriangledown_{\tau} n_1)_{\mu}\right)
\end{equation}
where $n_1=\partial_t$ and $n_2=\partial_x$, with $t$ and $x$ the boundary coordinates. The coefficient $\mu$ entering (\ref{nfun}) is connected to the boundary difference between central charges as
\[
\frac{1}{\mu}=\frac{c_R - c_L}{6} G_N.
\]

There are more solutions for $\Sigma$ than just geodesics, but geodesics are always a solution; similarly metric solutions to the TMG's equations of motion are broader than proper Einstein gravity solutions, but these are always solutions of TMG. As the analytic continuation is not clear in the broadest context, we will restrict as in \cite{Castro:2014tta} to usual geodesics in Einstein metrics. 

If we try to prove SSA2 as we have done for static spacetimes and general intervals in section \ref{faec} and repeat the steps leading to (\ref{sss}), we immediately face a complication: even requiring the integrated NCC (\ref{intncc}), the Raychaudhuri equation, that allowed us to minimize the proper length along the null geodesics flow of the congruence, does not constrain anymore the growth of the full functional (\ref{nfun}) as it is no longer simply the proper length of the curve $\Sigma$. Even try to enforce some modified energy condition to minimize (\ref{nfun}) along the null congruence flow and so keep SSA2 valid appears difficult. This because the term proportional to $1/\mu$ is essentially a boundary contribution. We can speculate this being a sign of the essential difference between a violation of SSA2 due to using unphysical backgrounds, and thus entirely avoidable by appropriate bulk conditions, and a violation due to a pure CFT mechanism with its dual holographic description.

Finally let me discuss briefly on SSA1. As in the static case we have here that SSA1 remains valid; the point is that SSA1 is generically a weaker condition than SSA2. In the static case we showed that its bound on the EE did not change by varying the interval configuration, remaining monotonicity the condition on $S(l)$. Here the EE is no longer a function of the sole proper length of the interval, and the discussion and the proof do not apply any longer. Still the moral appears to be the same, and an interesting development would be to try to generalize them to the present case of $c_L\neq c_R$ two dimensional CFTs.

\section{Time dependent case}

It is time to study time dependent situations, that for the CFT means a time dependent quantum state $\rho$ while for the bulk a time dependent metric. For what concerns the boundary side we unfortunately have not much to say that does not come from holography. The reason being first the missing knowledge of the state for computable backgrounds ( as the Vaidya example we will soon introduce ), and second the non invariance by time translation that spoils the usual dependence of the EE by only the proper length of the interval. This means that any EE computation, attempting to verify SSA1 and SSA2, should pass through the understanding of the bulk description. This we will now discuss in much more detail.

\subsection{Intervals at fixed boundary time}

When a time dependent background is considered, additional complications emerge in understanding the validity or not of strong subadditivity. First of all HRT surfaces $\Sigma(A)$ bend in the time direction and thus, even for collinear intervals, intersection does not generically happen. Second, inside the formula (\ref{ryutak}), $\Sigma(A)$ does not refer any longer to minimal surfaces as in the Euclidean static case, but to extremal ones. Consequently, even if we had intersection, we could not straightforwardly construct area inequalities as so far done ( unless we could " project " the extremal surfaces $\Sigma(A\cup B)$ and $\Sigma(B\cup C)$ to a certain common achronal slice containing $\Sigma(B)$ and $\Sigma(A\cup B\cup C)$, where they become minimal. We will soon see that non existence of this slice is the key point for failure of SSA2.).

A first understanding on how and when SSA1 and SSA2 are valid comes from using formula (\ref{r0st}), that is the specific case of (\ref{r0}) when the boundary interval and its variation are at fixed time. In this case, as we have previously derived,  (\ref{r0st}) is valid even without requirement of time translation invariance. Further in this case we still can write the EE as a a function of the interval length $\Delta x$, $S(\Delta x)$, with the dependence on the boundary time $t_b$ decoupling when studying inequalities at the same $t_b$ ( while for generic space-like intervals, without time translation invariance, the EE is a generic  function of all the coordinate end points, not only the proper length ). The positivity of the conformal factor at the vertex $r_0$ is just the monotonicity property that guarantees the validity of SSA1, but what about SSA2? Clearly to have the EE $S(\Delta x)$ concave as a function of $\Delta x$, we should have that $r_0(\Delta x)$  ( at fixed time $t_b$  ) is monotonically decreasing or, by inverting the function for given $t_b$, that $\Delta x(r_0)$ is monotonically decreasing  \footnote{Remember, to avoid confusion, that big $r$ is close to the boundary and small $r$ to the center of the bulk.}. This means that violation of concavity and SSA2 happens for geodesics whose vertex moves towards the boundary when extending the size $\Delta x$ of the interval, for fixed $t_b$. 

To understand better this point we can use a theorem by Wall, \cite{Wall:2012uf} theorem 17, that says that if NCC is valid, two geodesics ( maximin surfaces in the paper as we will discuss later ) ending on space-like boundaries one contained in the other, $A(\Delta x,t_b)$ and $A(\Delta x+\delta x,t_b)$ in our case, will always be at space-like distance one from the other with the smaller one inside, towards the boundary, with respect to the bigger. So SSA2 is respected. Instead giving up NCC not only time-like distances are possible but, in order to have violation of SSA2 the narrower geodesic should have its vertex extending in the bulk further then the larger one.

\subsection{Vaidya example for fixed time intervals} 
 
To be more concrete let us consider an example of time dependent, asymptotically AdS background, where analytic computations are possible. This is the Vaidya metric, representing the collapse of a mass shell that interpolates between an AdS metric and a BTZ black hole. This example has been extensively studied in the past by  \cite{Allais:2011ys}, \cite{Caceres:2013dma} and \cite{Callan:2012ip}, where the first connection between the NCC and the violation of SSA2 was established. 

The metric is 
\begin{equation}
ds^2=-(r^2 -m(v))dv^2 +2 dr dv + r^2 dx^2
\end{equation}
which is an AdS solution with the addition of a local energy momentum tensor whose only nonzero component is 
\begin{equation}
T_{vv}=\frac{1}{2r}\partial_v m(v).
\end{equation}
The case we consider is when $m(v)$ is a step function on $v=0$ and $T_{vv}$ becomes a delta. If the delta is positive ( $m(v<0)=0,\;m(v>0)=m$ ) the metric will be AdS inside for $v<0$ and  BTZ outside for $v>0$; if instead the delta is negative  ( $m(v<0)=m,\;m(v>0)=0$ ) we have BTZ inside for $v<0$ and AdS outside for $v>0$. The first case satisfies the NCC, the second violates it, as can be trivially checked. From now on $m=1$ \footnote{ There is an associated scaling symmetry that allows this choice, see for example \cite{Callan:2012ip}. }.

\begin{figure}[h]
\centering
\vspace{-10pt}
\includegraphics[width=1\textwidth]{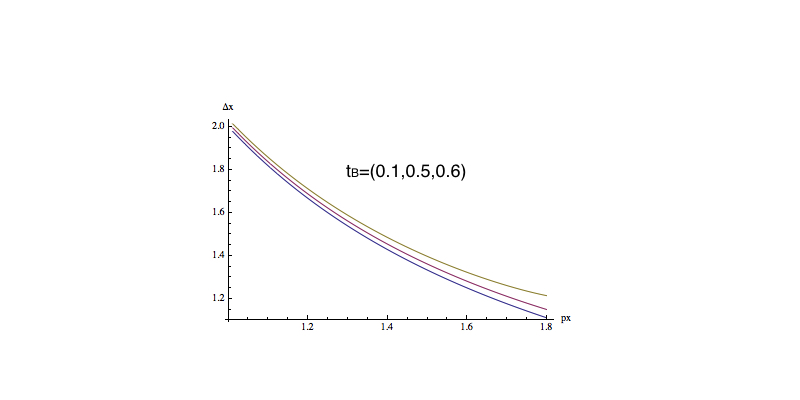}
\vspace{-30pt}
\caption{SSA2 respected by a monotonically decreasing behaviour of $\Delta x(p_x)$, for Vaidya background respecting NCC.}
\label{figuratra2}
\end{figure}

Geodesics that starts and ends on the boundary and cross the mass shell ( otherwise they are simply contained in the static AdS or BTZ depending on the choice for $m(v)$ ) depend on two parameters that, following \cite{Callan:2012ip},  we choose to be $r_c$ and $p_x$, the bulk radius at which the geodesic crosses the shell and a conserved momentum for the space translation symmetry that turns out to correspond to the radius of the vertex. We start with backgrounds respecting NCC. The goal is to verify the monotonically decreasing behaviour of $\Delta x(r_0)$, where $r_0=p_x$ as explained in appendix \ref{B} and where relevant formulas can be found; $\Delta x$ is there called $\Delta x_{b}$ and is a function of  $r_c$ and $p_x$. We solve for $r_c$ to give, for the geodesic with a certain value of $p_x$, the chosen boundary time $t_b$, and then plot $\Delta x(p_x)$ for the given value of $t_b$. Some sample curves are represented in figure \ref{figuratra2} where we can check the monotonically decreasing behaviour, as expected.
More interesting is the case for NCC violating geometries. Here formulas depend on the range of value for the parameters, with three possible cases. Case 1: $r_c > p_x > 1$; case 2: $1> r_c > \sqrt{1/2} $ and $1>p_x > r_c $; case 3: $\sqrt{1/2}> r_c >0$ and $p_x^2 - E_A^2 > 0$ with $E_A$ a certain function of $r_c$ and $p_x$ whose formula and meaning is in appendix \ref{B}. Some sample plots are in figure \ref{figuratra}, where we can see both monotonically decreasing curves, for geodesics belonging to case 1 and thus respecting SSA2, and monotonically increasing for geodesics in cases 2 and 3, so violating SSA2. 
\begin{figure}[h]
\centering
\vspace{-10pt}
\includegraphics[width=1.2\textwidth]{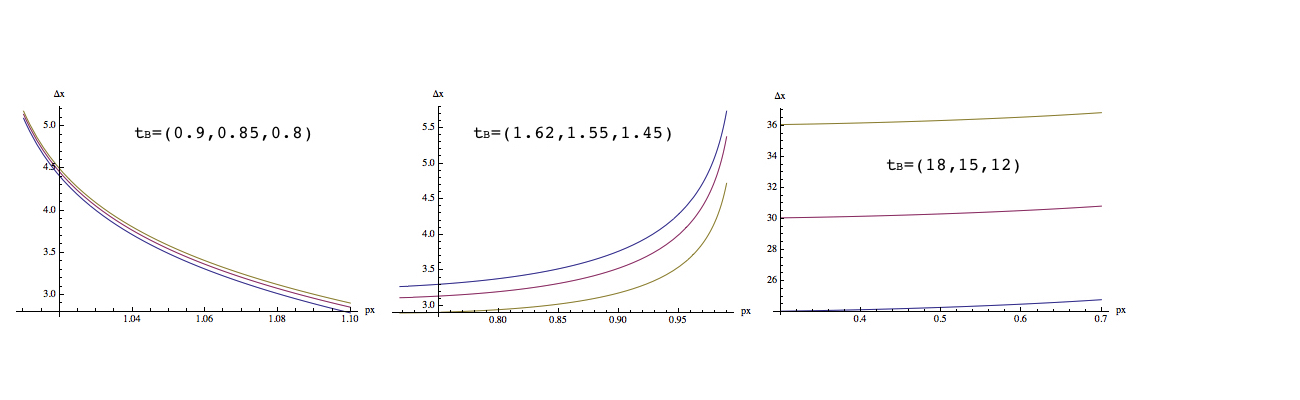}
\vspace{-30pt}
\caption{SSA2 respected and violated by monotonically decreasing and increasing function $\Delta x(p_x)$, for Vaidya background violating NCC. The pictures refer to case 1, case 2 and case 3 respectively. }
\label{figuratra}
\end{figure}

\subsection{Generic space-like intervals}

Let us now discuss the most general scenario, with time dependent backgrounds and generic adjacent interval configurations \footnote{This section greatly benefited by e-mail correspondence with Aron C. Wall and Horacio Casini.}. The starting point for the discussion is the  paper by Wall \cite{Wall:2012uf}, where the author introduces the concept of  maximin surfaces, as an equivalent description of the extremal surfaces $\Sigma(A)$. A codimension two maximin surface is defined starting with a generic codimension one achronal surface $T$ in the bulk containing the boundary of the interval $A$; then a maximin surface is the minimal codimension one surface on $T$ having maximal area when varied over all the possible $T$. In brief the extremality is achieved by minimizing and maximizing in the space and time directions respectively. 

\begin{figure}[h]
\centering
\vspace{-0pt}
\includegraphics[width=0.9\textwidth]{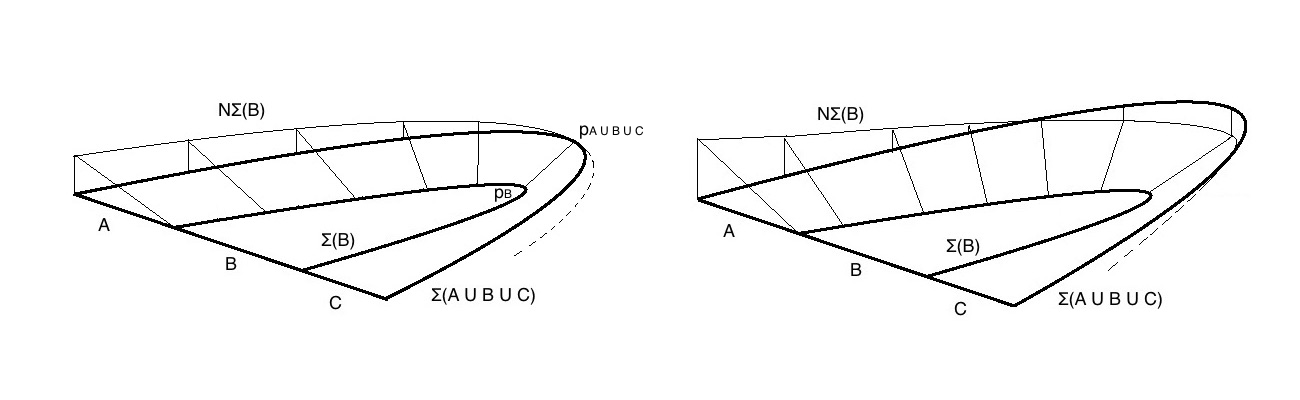}
\vspace{-0pt}
\caption{Developement of time-like distances between the two extremal surfaces $\Sigma{B}$ and $\Sigma(A\cup B\cup C)$.}
\label{figuratld}
\end{figure}

The power of the maximin construction is exploited in the theorem that states ( under generic assumptions for the bulk spacetime ) that SSA2 is valid if the NCC condition
\begin{equation}\label{ncc}
R_{\mu\nu}k^{\mu}k^{\nu}\geq 0
\end{equation}
is respected. The original proof in \cite{Wall:2012uf} uses NCC and the Raychaudhuri equation for constructing inequalities between areas on different achronal slices with the same boundary condition. This result may be obtained as well just by using the maximin construction alone, so in order to single out the places where the NCC necessarily enters as a condition for SSA2, and to be naturally lead to the main claim of this section, we slightly modify the proof. Skipping some details, that may be found in the original paper, it is:

Proof : theorem 4 of \cite{Wall:2012uf} states that, given two null congruences of geodesics $N_1$ and $N_2$, in our case obtained by shooting out null curves from maximin surfaces, with $N_2$ nowhere in the past of $N_1$ and touching at a point $p$ belonging to some achronal slice $T$, then there exists a sufficiently small neighbourhood of $p$, $B_p\in T$, such that either $\Theta(N_2)_{B_p}>\Theta(N_1)_{B_p}$ or $N_1$ and $N_2$ do coincide there. Given this general result, theorem 14 states that two maximin surfaces with space-like boundary condition ( attached to $\Sigma(B)$ and $\Sigma(A\cup B\cup C)$ in our case ), are always at space-like distance if NCC holds; the idea is simply that, as maximin surfaces are extremal, on $\Sigma(B)$ and $\Sigma(A\cup B\cup C)$ we have $\Theta(N(\Sigma(B)))_{\Sigma(B)}=\Theta(N(\Sigma(A\cup B\cup C)))_{\Sigma(A\cup B\cup C)}=0$; then starting from a situation where the two curves are always at a space-like distance between them, not only close to the boundary but all the way through the bulk, let us suppose we can continuously deform the curves, for example enlarging $A\cup B\cup C$ while keeping $B$ fixed, to a situation where somewhere the proper distance approaches the null value; this means that two points $p_B$ and $p_{A\cup B\cup C}$, one for each curve, are connected by a null geodesics ( for symmetry there is either a single $p$ corresponding to the vertex or two symmetric points on the right and left hand side of the vertex ). This is shown in the first picture of figure \ref{figuratld}.
Let us choose  $\Sigma(A\cup B\cup C)$ to be nowhere in the past of $\Sigma(B)$ and focus on $p_B$; theorem 4 says that in a neighbourhood of $p_B$ the null congruence $\Theta(N(\Sigma(A\cup B\cup C)))_{B_{p_B}}>\Theta(N(\Sigma(B)))_{B_{p_B}}$. As NCC is valid, $\Theta(N(\Sigma(A\cup B\cup C)))_{\Sigma(A\cup B\cup C)}>\Theta(N(\Sigma(A\cup B\cup C)))_{B_{p_B}}>\Theta(N(\Sigma(B)))_{B_{p_B}}=0$ and this contradicts the extremality condition. Not being able to continuously approach a light-like distance between any two points, it is then implied that the two curves cannot be deformed to develop time-like distances either, as shown in the second picture of figure \ref{figuratld}. This point is crucial for proving theorem 17 that shows that both $\Sigma(B)$ and $\Sigma(A\cup B\cup C)$ always belong to the same achronal slice $T$. Finally the key passage is that, using the maximin construction, there exists a representative for both $\Sigma(A\cup B)$ and $\Sigma(B\cup C)$ on $T$ of smaller area ( proper length ) then the maximin surfaces, $\mathcal{A}(\Sigma(A\cup B))>\mathcal{A}(\Sigma(A\cup B)_T)$ ( and analogously for  $\Sigma(B\cup C)$ )\footnote{This is the point where we do not make use of NCC that is instead used by \cite{Wall:2012uf} to compare areas.}. Thus on $T$ we have 
\begin{equation}
\mathcal{A}(\Sigma(A\cup B))+\mathcal{A}(\Sigma(B\cup C))>\mathcal{A}(\Sigma(A\cup B)_T)+\mathcal{A}(\Sigma(B\cup C)_T)\geq \mathcal{A}(\Sigma(B))+\mathcal{A}(\Sigma(A\cup B\cup C))
\end{equation}
where the last inequality is just the usual static argument ( that can now be used as on $T$ the surfaces $\Sigma(B)$ and $\Sigma(A\cup B\cup C)$ are minimal and $\Sigma(B)_T$ and $\Sigma(A\cup B\cup C)_T$ intersects. ).

This formulation of the theorem makes evident that a necessary condition for violation of SSA2 is to develop non space-like distances between $\Sigma(B)$ and $\Sigma(A\cup B\cup C)$. This is understood as NCC  enters the above proof only once, in constraining $\Sigma(B)$ and $\Sigma(A\cup B\cup C)$ to belong to the same achronal slice $T$ while remaining at space-like distances. 

The importance of this result, that is the non space-like distance between two surfaces $\Sigma(A_1)$ and $\Sigma(A_1)$, with the domain of dependence of $A_1$, $D_{A_1}$, containing $D_{A_2}$, $D_{A_1}\in D_{A_2}$, resides in being a counter example to statements sometimes used in the past literature, for instance Conjecture C2 of \cite{Czech:2012bh}.

We would like to emphasize the difference between the local NCC energy condition obtained in the present section, and the integrated NCC that we obtained for static spacetimes. The former one is clearly more restricting than the second, as respecting the local NCC obviously implies the integrated NCC, but not the opposite. In fact we could have used the maximin construction for proving that local NCC implies SSA2 for static backgrounds ( which is true ), but the maximin construction requires the local NCC to be applicable ( for example in proving the equivalence with HRT surfaces ). So in order to have the theorem as strong as possible, by requiring the weakest energy condition, we proceeded there without this powerful tool.  

\subsection{Vaidya example for generic space-like intervals} 

Given the generic discussion of the past section, let us construct a concrete example, where we show the developing of either null or time-like distances between the two geodesics $\Sigma(B)$ and $\Sigma(A\cup B\cup C)$, when NCC does not hold and SSA2 is violated, while in general maintaining space-like distances when NCC is valid ( although it is possible to have SSA2 respected in the former case as the condition is necessary but not sufficient ).
In this section we will study only situations where both the $\Sigma(B)$ and $\Sigma(A\cup B\cup C)$ geodesics belong to the same parameter range ( case 1,2,3 as previously introduced, see also appendix \ref{B} ), even though in appendix \ref{B} formulae are provided for the most general scenario. 

Our goal will be to probe the distance between the vertices of $\Sigma(B)$ and $\Sigma(A\cup B\cup C)$ to check if they are at space-like distance or not, depending on the value of the parameters $r_c$ and $p_x$ of both curves. We will not consider distances between generic points on the geodesics as it would be excessively complicated to derive corresponding inequalities and ultimately unnecessary, as non space-like distances between the two curves is just a necessary but not sufficient condition for SSA2 violation. As the boundary conditions force space-like distance between the end points of $\Sigma(B)$ and $\Sigma(A\cup B\cup C)$, the vertices are likely to be the most prone to develop either null or time-like distances between them, so we restrict to this case. Further to simplify an otherwise too complicated computation we will consider only collinear intervals at the same boundary time $t_b$ and symmetrically disposed around the central point of $B$, that is $A$ and $C$ are taken to be of the same length.

In appendix \ref{B} we derived inequalities for the parameters $r_c$ and $p_x$ of two geodesics $\Sigma(B)$ and $\Sigma(A\cup B\cup C)$ that, when respected, correspond to space-like distances between the vertices. The constrain that comes from having both curves end points at the same boundary time $t_b$ eliminates one of the four parameters and further partially restricts the available parameter space that comes from the other three. The results are the following. \\

\textbf{Metric that violates the NCC}: \\

We here generically name "1" and "2" the labels for the two geodesics attached to the two parameters $r_c$ and $p_x$. As the goal is just to show if the distance between the vertices is space-like or not, it is not really relevant to distinguish which one corresponds to the geodesic attached to the largest boundary interval $\Sigma(A\cup B\cup C)$, and which to $\Sigma(B)$. However, as the intervals are collinear, we know that for case 1 smaller value of $p_x$ corresponds to bigger value for $\Delta_x$, so that parameter belong to $\Sigma(A\cup B\cup C)$. For case 2 and 3 instead, the smaller $p_x$ belong to $\Sigma(B)$.

\begin{itemize}
\item{Case 2: $1> r_c > \sqrt{1/2} ,\;1>p_x > r_c $

The inequalities defining space-like distance between the vertices are ( see appendix \ref{B} ):
 \begin{subequations}
\begin{align}
y(p_{x1}) y(r_{c2}) &< y(p_{x2}) y(r_{c1})  \label{in21a}\\
& or \nonumber \\
y(r_{c1}) y(p_{x1}) &< y(p_{x2}) y(r_{c2})\label{in22b} 
\end{align}
\end{subequations}
where we have defined
\begin{equation}
y(r)=\frac{1 + r}{-1 + r} 
\end{equation}
in a parameter space spanned by $1>p_{x1}>r_{c1}>r_{c2}> \sqrt{1/2}$ with $r_{c1}>r_{c2}$ and $p_{x1}>r_{c1}$. $p_{x2}$ is a function of the other three parameters, as explained in appendix \ref{B}. We can sweep this parameter space to look for what volume satisfies the inequalities (\ref{in21a}) and  (\ref{in22b}) and what violates them. The result is that they are always violated as shown in figure \ref{figura5} and thus geodesics always develop non space-like distances. }
\begin{figure}[h]
\centering
\vspace{+10pt}
\includegraphics[width=0.8\textwidth]{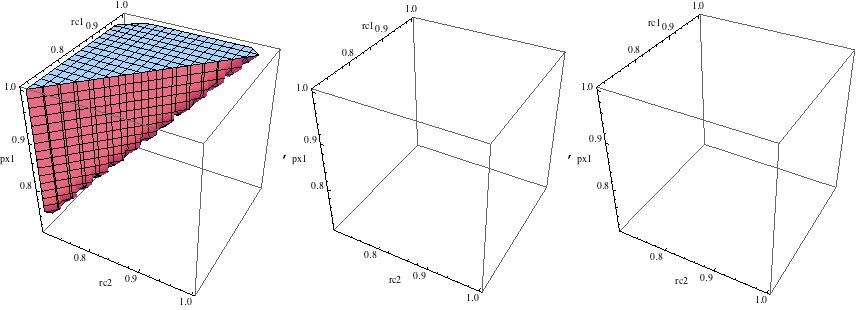}
\vspace{-0pt}
\caption{The first graph shows the volume in the parameter space that violates both (\ref{in21a}) and  (\ref{in22b}), thus giving null or time-like distances between the vertices; the second and third show the ( empty ) region that satisfies (\ref{in21a}) and  (\ref{in22b}) respectively, thus giving space-like distances. The small angle missing is due to the requirement $t_b(r_{c1},p_{x1})-t_b(r_{c2},p_{x2})=0$ . }
\label{figura5}
\end{figure}

\item{Case 3: $\sqrt{1/2}> r_c >0, \;p_x^2 - E_A^2 > 0$

The inequalities are as previously:
 \begin{subequations}
\begin{align}
y(p_{x1}) y(r_{c2}) &< y(p_{x2}) y(r_{c1})  \label{in31a}\\
& or \nonumber \\
y(r_{c1}) y(p_{x1}) &< y(p_{x2}) y(r_{c2}).\label{in32b} 
\end{align}
\end{subequations}
inside a parameter space $p_{x1}>r_{c1}>r_{c2}$ with the additional constrain $p_{x1} < r_{c1}/(1 - 2 r_{c1}^2)$ and again $p_{x2}$ being a function of the other three parameters. Again always violation, as shown in figure \ref{figura5b} }
\begin{figure}[h]
\centering
\vspace{-0pt}
\includegraphics[width=0.8\textwidth]{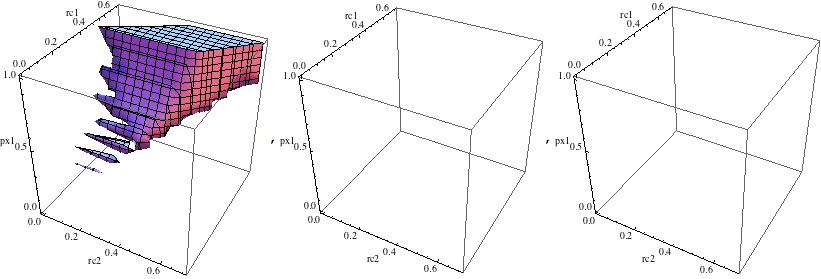}
\vspace{-0pt}
\caption{Graphs showing parameters region giving space-like ( first ) and null or time-like distances ( second and third, both empty ) between the vertices of the two geodesics. Relevant inequalities are (\ref{in31a}) and (\ref{in32b}). Note that the parameter space is different from the previous section not only in the values of the domains for $r_{c1},p_{x1},r_{c2}$ but also in shape. }
\label{figura5b}
\end{figure}

\item{Case 1: $r_c > p_x > 1$

Inequalities here are slightly different:
\begin{subequations}
\begin{align}
y(r_{c1}) y(p_{x1}) &> y(p_{x2}) y(r_{c2})  \;\;\;and \;\;\; y(p_{x1}) y(r_{c2}) > y(p_{x2}) y(r_{c1})  \;\;\;p_{x1}<p_{x2}\label{in11a}\\
y(r_{c2}) y(p_{x2}) &> y(p_{x1}) y(r_{c1}) \;\;\;and \;\;\; y(p_{x2}) y(r_{c1}) > y(p_{x1}) y(r_{c2})  \;\;\;p_{x1}>p_{x2}.\label{in12b}
\end{align}
\end{subequations}
with the parameter range for $r_{c1},p_{x1},r_{c2}$  limited by the constraints $r_{c1}>p_{x1}$ and  $r_{c2}>p_{x2}(r_{c1},p_{x1},r_{c2})>1$, and  $p_{x2}$ again a function to match the boundary time $t_b$. Inside this volume we separately deal with the subspaces $p_{x1}>p_{x2}$ and $p_{x1}<p_{x2}$; Now we can apply (\ref{in11a})  and (\ref{in12b}) on the relevant space and see if they are satisfied or not. It turns out that when  $p_{x1}<p_{x2}$, $y(p_{x1}) y(r_{c2}) > y(p_{x2}) y(r_{c1}) $ is always respected, while $y(r_{c1}) y(p_{x1}) > y(p_{x2}) y(r_{c2})$ is respected in some region and violated in its complementary, as is shown in figure \ref{figura7}. Thus here we face geodesics that both present vertices at space-like and non space-like distances, depending on the values of their parameters. 
\begin{figure}[h]
\centering
\vspace{+10pt}
\includegraphics[width=0.6\textwidth]{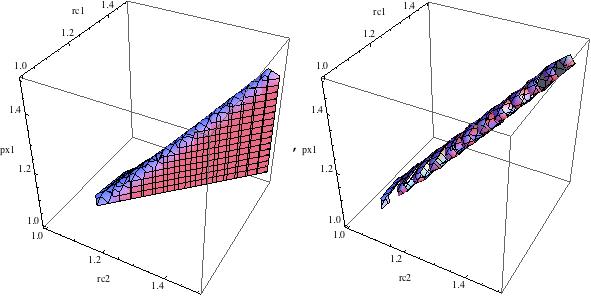}
\vspace{-0pt}
\caption{$p_{x1}<p_{x2}$ case: agreement and violation of $y(r_{c1}) y(p_{x1}) > y(p_{x2}) y(r_{c2})$ divide the total parameter space in two complementary subspaces. As the other inequality of (\ref{in11a}) is always respected, we have space-like distances in the first case, and null or time-like in the second.}
\label{figura7}
\end{figure}
Correspondingly when $p_{x1}>p_{x2}$ $y(r_{c2}) y(p_{x2}) > y(p_{x1}) y(r_{c1})$ is always respected, while $y(p_{x2}) y(r_{c1}) > y(p_{x1}) y(r_{c2}) $ is respected in some region and violated in its complementary ( with respect to the total space  ). This is shown in figure \ref{figura8}.}

\begin{figure}[h]
\centering
\vspace{-0pt}
\includegraphics[width=0.6\textwidth]{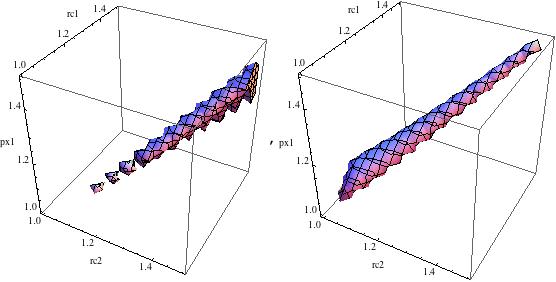}
\vspace{-0pt}
\caption{$p_{x1}>p_{x2}$: agreement and violation of  $y(p_{x2}) y(r_{c1}) > y(p_{x1}) y(r_{c2}) $ divide the total parameter space in two complementary subspaces. As the other inequality of (\ref{in12b}) is always respected, we have space-like distances in the first case, and null or time-like in the second.}
\label{figura8}
\end{figure}
\end{itemize}

\textbf{Metric that respects the NCC}: \\

The space-like condition for distances between vertices is here given by the inequalities 
\begin{subequations}
\begin{align}
\frac{1}{p_{x1}} + \frac{1}{r_{c1}}>\frac{1}{p_{x2}} + \frac{1}{r_{c2}}  \;\;\;\; and \;\;\;\; \frac{1}{p_{x1}} + \frac{1}{r_{c2}}>\frac{1}{p_{x2}} + \frac{1}{r_{c1}} \;\;\;p_{x1}<p_{x2} \label{csl1} \\
\frac{1}{p_{x1}} + \frac{1}{r_{c1}}<\frac{1}{p_{x2}} + \frac{1}{r_{c2}}  \;\;\;\; and \;\;\;\; \frac{1}{p_{x2}} + \frac{1}{r_{c1}}>\frac{1}{p_{x1}} + \frac{1}{r_{c2}} \;\;\;p_{x1}>p_{x2} \label{csl2} 
\end{align}
\end{subequations}
with only $r_{c1},p_{x1},r_{c2}$ independent and $p_{x1}<p_{x2}$. The parameter space defined by the usual condition for the boundary time and either (\ref{csl1}) or (\ref{csl2}) may be obtained numerically, in chosen parameter domains. The result is that (\ref{csl1}) and (\ref{csl2}) are always respected implying vertices are at space-like distance one from the other.

\section{Conclusions}

We have discussed in detail the holographic description of the two strong subadditivity inequalities, from static backgrounds to time dependent, with collinear boundary intervals or general configurations. We have seen that, for static geometries, SSA1 and SSA2 are always respected for collinear intervals, that the second requires an integrated NCC for non collinear intervals and is generically violated when we have Lorentz anomaly, while SSA1 holds. This has its counterpart in the monotonicity condition for SSA1 that remains unaltered, while the concavity for SSA2 transforms into a more strict requirement abandoning the collinearity. New results are the geometric proof for concavity of minimal surfaces, the proof that SSA1 implies only monotonicity independent of the interval configuration, a new proof that SSA2 requires the integrated NCC using the Raychaudhuri equation and finally the violation of SSA2 but not of SSA1 ( here only numerical ) for Lorentz anomalous CFTs. For time dependent backgrounds we first provided a new simple strategy for understanding if and when SSA1 and SSA2 violation occurs with collinear intervals, that does not requires direct checking of monotonicity or concavity. Second we have reviewed, in a slightly different form, the result by Wall that local NCC implies validity for SSA2, while making manifest the connection with the energy condition by showing that the reason for the violation comes from the developing of null or time-like distances between the most inward and outward geodesics entering the SSA2 inequality. Furthermore we have provided an explicit example in both cases by using the Vaidya metric. Also some discussion on why violation of strong subadditivity occurs has been provided. Following are two appendixes containing the results and the proofs ( to my knowledge new ) for what interval configuration, as a function of the slopes, gives the strongest bound on the entanglement entropy inequalities SSA1 and SSA2, and explicit formulas ( and some derivation ) for the Vaidya metric example.

An interesting point of view, that we did not discuss but is worth mentioning, is the result from  \cite{Parikh:2014mja} where it was shown that Virasoro conditions in bosonic string theory imply the NCC on the background geometry. We would like to suggest that, perhaps, this is a hint that energy conditions may have some UV justification. The hypothesis is that non respecting NCC ( or analogous conditions ) means that the metric we are using does not consistently arise as a background in theories that correctly quantize gravity, and one of the dual symptoms is not respecting strong subadditivity.

We would like to point out three possible hints for future research. The first one is the problem of quantum bulk corrections to EE and the question if they do respect or not the boundary strong subadditivity inequality. For example we can introduce the mutual information $I(A,B)\equiv S(A)+S(B)-S(A\cup B)$ and rewrite the SSA2 inequality as, \cite{Allais:2011ys}
\begin{equation}\label{inmut}
I(A,B\cup C)\geq I(A,B).
\end{equation}
If the intervals $A$ and $B$ entering the mutual information are well disconnected, the classic holographic computation gives $I(A,B)=0$, as $\Sigma(A\cup B)=\Sigma(A)\cup \Sigma(B)$. Thus, for $A$, $B$ and $C$ disconnected the inequality (\ref{inmut}) just produces a classical $0=0$ result. Quantum corrections in the bulk clearly affect the above inequality; how to compute them is still a discussed argument as the HRT ( or Ryu-Takayanagi ) surfaces are not string worldsheets as for the holographic description of Wilson loops, but rather just geometrical surfaces. Thus $\alpha'$ corrections or higher genus computations are not the correct answer. The question has so far received two different answers in \cite{Engelhardt:2014gca} and \cite{Faulkner:2013ana}, with pro and con arguments for both. The question is then if  there are reasons to believe that quantum bulk corrections do respect or not strong subadditivity, and/or should they be constrained by energy conditions in doing so? 

A second direction for future research, and also one of the original motivations for the present paper, is the question of what is the logical relationship between strong subadditivity and energy conditions. We know that the integrated NCC implies SSA2 for static geometries with generic adjacent intervals, and that the two are equivalent for a SSA2-bound-maximizing configuration, and we also know that local NCC implies SSA2 in time dependent systems. Can we deform either the energy condition or the strong subadditivity, weakening or strengthening depending on the case, in order to make the correspondence one to one, in the widest possible range of cases? 

Finally we would like to generalize as most as possible of the present paper to generic d-dimensional theories. Part of this work is straightforward, part quite complicated.

We hope to come back to these issues and more in the future.

\section*{Acknowledgements}

I would like to thank Aron C. Wall and Horacio Casini for email correspondence at the early stage of this work, and Diego Trancanelli for suggestions and comments on the paper. This work was founded by FAPESP fellowship 2013/10460-9.

\appendix
\section{Most constraining interval configurations, proof for the two inequalities}\label{A}
In this appendix we want to study what are the configurations of one dimensional adjacent connected intervals $A,B,C$ such that the EE is maximally constrained by the inequalities of strong subadditivity (\ref{ssa1}) and (\ref{ssa2}). 

As explained in \cite{Casini:2004bw}, in theories that respect causality and unitarity, the dependence on the space-like interval $A$ of $S(A)$ can be only through the causal domain of dependence of $A$ that, in two dimensions ( without global issues ), is determined exclusively by the two endpoints of the interval; if the theory has Lorentz invariance the dependence should be further restricted to the proper distance between them. This is the only assumption we will make on $S(l(A))$, that does not come from strong subadditivity. 

The requirement for $A,B,C$ to be space-like ( or light-like in some appropriate limit ) may, at first sight, appear excessive; we can, for example, consider time-like $A$ and/or $C$ while still having space-like separated end points of $A\cup B$ and $B\cup C$ ( and consequently also $A\cup B\cup C$ ), and an apparently meaningful SSA2 inequality (\ref{ssa2}) \footnote{Similarly we can pick a time-like interval $B$ with space-like end points for $A\cup B$ and $B\cup C$ and consider the SSA1 (\ref{ssa1}). }. One way to see that this is not possible is from the explicit proof of SSA1 and SSA2 \cite{Chuang}, where Hilbert spaces associated to all the three $A,B$ and $C$ are involved. Another way is to show contradiction for (\ref{ssa2}).
 \begin{figure}[h]
\centering
\vspace{-0pt}
\includegraphics[width=0.7\textwidth]{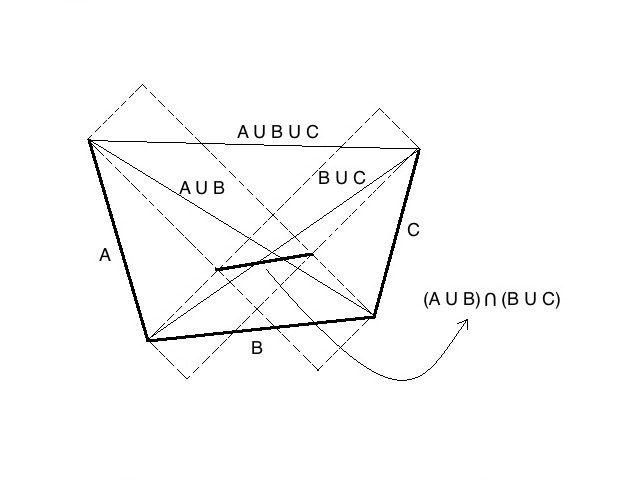}
\vspace{-40pt}
\caption{Intervals $A$ and $C$ are time-like such that the end points of $A\cup B$ and $B\cup C$ are at barely space-like distances. Note that in this example the intersection between the causal domains of $A\cup B$ and $B\cup C$ determine a region whose most outwards end points in the space direction fix what we call $(A\cup B) \cap (B\cup C)$; this segment does not coincide with $B$, as it should be by causality and would indeed happen if $A$ and $C$ were space-like. Violation of SSA2 occurs. }
\label{figuraa}
\end{figure}
Consider the picture \ref{figuraa} where $A$ and $C$ are time-like such that $A\cup B$ and $B\cup C$ are barely space-like, that is with positive proper distance tending to zero. The intersection between their domains of dependence should determine the space-like segment $B$, but it clearly does not. If one still tries to use (\ref{ssa2}), written to respect causality as $S(A\cup B) + S(B\cup C)\geq S((A\cup B) \cap (B\cup C))+S(A\cup B \cup C)$, ends up with an inequality where the proper lengths  $l(A\cup B)$, $l( B\cup C)$ and $l((A\cup B) \cap (B\cup C))$ can be made arbitrary small while $l(A\cup B \cup C)$ in general remains finite.
As we know that $S(l(A))$ is a monotonically increasing function, this clearly leads to a contradiction. 

\begin{figure}[h]
\centering
\vspace{-0pt}
\includegraphics[width=1.0\textwidth]{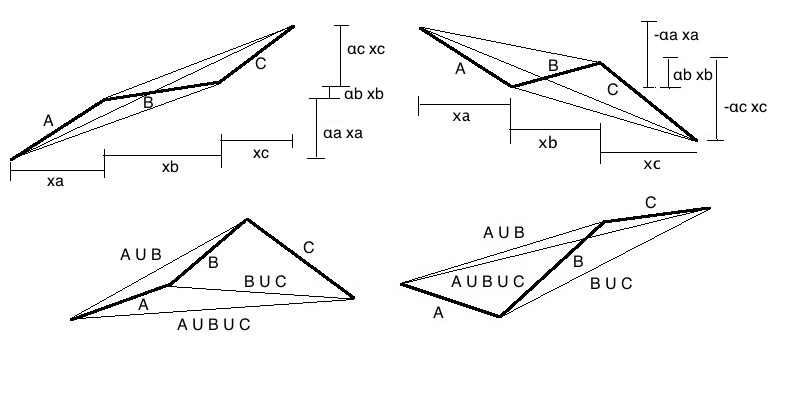}
\vspace{-40pt}
\caption{PPP NPN PPN NPP configurations of intervals, showing parameterization in the first two, and straight space-like segments associated with composite intervals in the last two. }
\label{figuracldcr}
\end{figure}

Let us label the different possible configurations of intervals $A,B,C$ by the signs of their slopes: $P$ for positive and $N$ for negative. We have eight cases that, assuming parity invariance for the space $x$-coordinate direction, are related as below 
\begin{equation}\label{ppp}
(PPP, PPN, NPP, NPN) \xleftrightarrow{\;\;\;P\;\;\;} (NNN, PNN, NNP, PNP).
\end{equation}
As in figure \ref{figuracldcr} we call $x_A, x_B,x_C$ and $\alpha_A x_A,\alpha_B x_B,\alpha_C x_C$ the space $x$ and time $t$ coordinate distances, with $\alpha$ the slope defined by the counterclockwise angle measured from the $x$-axis ( as usual ), $x_A,x_B,x_C>0$, $1\geq\alpha_{A,C}\geq-1$, $\alpha_B \geq 0$. Time and space distances of composed intervals, for instance $A\cup B$, will be the sum of the corresponding ones for $A$ and $B$; we will call slope of any interval the ratio between the time and space total distances. The case $\alpha_A=\alpha_B=\alpha_C=0$ has been already considered in section \ref{cft}, to fix $S(l)$ to be monotonically increasing and concave, so these properties will be assumed. Further we will work in Lorentzian signature so, for instance $l(A)=\sqrt{x_A^2-(\alpha_Ax_A)^2}$ etc...

The general strategy is to increase ( resp. decrease ) as much as possible the right-hand ( resp. left-hand )  sides of (\ref{ssa1}) and (\ref{ssa2}), by increasing ( resp. decreasing ) the proper lengths of the intervals \footnote{Remember that $S(l)$ is monotonically increasing.} by varying $\alpha_A,\alpha_B,\alpha_C$ while leaving $x_A,x_B,x_C$ generic ( they will be adjusted in some cases ).  

If a variation of some parameter $\alpha$ acts in such a way to contemporary grow the smallest side of the inequality ( either SSA1 or SSA2 ) and reduce the largest, then this variation is applied as it makes the bound from that inequality on the function  $S(l)$ stronger. Instead
it may happen that changing some $\alpha$'s leads to either increasing or decreasing both sides of the inequality at the same time. Still, using the fact that $S(l)$ is monotonically increasing and concave, if we have two intervals with proper lengths $l_1$ and $l_2$ depending on that $\alpha$, $l_2\geq l_1$, on opposite sides of either SSA1 or SSA2, there are cases where, without knowing the exact form of the function $S(l)$, we can still say if $S(l_2)-S(l_1)$ augments or diminishes:
\begin{subequations}
\begin{align}
\Delta(S(l_2)-S(l_1))>0  \;\;\; for\;\;\;( \Delta l_1 <0,\Delta l_2 >0) \;\;\; or\;\;\; ( \Delta l_1 <0,\Delta l_2 <0,\Big|\frac{\partial l_1(\alpha_1)}{\partial \alpha_1}\Big|\geq\Big|\frac{\partial l_2(\alpha_2)}{\partial \alpha_2}\Big|) \label{der1}  \\
 \Delta(S(l_2)-S(l_1))<0  \;\;\; for\;\;\;( \Delta l_1 >0,\Delta l_2 <0) \;\;\; or\;\;\; ( \Delta l_1 >0,\Delta l_2 >0,\Big|\frac{\partial l_1(\alpha_1)}{\partial \alpha_1}\Big|\geq\Big|\frac{\partial l_2(\alpha_2)}{\partial \alpha_2}\Big|). \label{der2}  
\end{align}
\end{subequations}
Every other variation of $l_1$ and $l_2$ leads to an undetermined sign for $S(l_2)-S(l_1)$. 

\subsection{SSA2}

Let us start with the second inequality (\ref{ssa2}) and consider the possible configurations listed in the left side of (\ref{ppp}). From any of the pictures of figure \ref{figuracldcr} we can immediately understand that minimizing $l(A\cup B)$ and at the same time maximizing $l(A\cup B\cup C)$ by varying $\alpha_A$ ( or equivalently with $A\leftrightarrow C$ ), is possible only if $(x_A \alpha_A + x_B \alpha_B + x_C \alpha_C) < 0$ and $(x_A \alpha_A + x_B \alpha_B) > 0$ ( increasing $\alpha_A$ ) or if $(x_A \alpha_A + x_B \alpha_B + x_C \alpha_C) \geq 0$ and $(x_A \alpha_A + x_B \alpha_B) \leq 0$ ( decreasing $\alpha_A$ ). If any of these conditions is fulfilled we obtain a stronger bound from SSA2 by changing $\alpha_A$ until it reaches its limiting value $\pm1$ or the value for which the condition itself is no longer valid; if $\alpha_A$ has reached a value different from $\pm1$, or if the above conditions were not fulfilled from the very beginning, the way to maximize $S(A\cup B\cup C)- S(A\cup B)$ is to compare derivatives, as on the right hand side of either (\ref{der1}) or (\ref{der2}). 

\subsubsection{PPN}

In the present case $(x_A \alpha_A + x_B \alpha_B) \geq 0$, thus if it is strictly positive and $(x_A\alpha_A + x_B \alpha_B + x_C \alpha_C) < 0$ we should increase $\alpha_A$ either to $+1$ or to the maximum value for which $A\cup B\cup C$ becomes flat. If $(x_A\alpha_A + x_B \alpha_B + x_C \alpha_C) \geq 0$ ( either as an initial condition or by effect of the growth of $\alpha_A$ ), then $\frac{\partial l(A\cup B)(\alpha_A)}{\partial \alpha_A}\leq 0$ and $\frac{\partial l(A\cup B\cup C)(\alpha_A)}{\partial \alpha_A}\leq 0$. The condition to maximize $S(A\cup B\cup C)- S(A\cup B)$ is $\Big|\frac{\partial l(A\cup B)(\alpha_A)}{\partial \alpha_A}\Big|\geq\Big|\frac{\partial l(A\cup B\cup C)(\alpha_A)}{\partial \alpha_A}\Big|$ that simple algebra transforms into
\begin{equation}\label{slope}
\frac{x_A \alpha_A + x_B \alpha_B}{x_A + x_B} \geq \frac{x_A \alpha_A + x_B \alpha_B + x_C \alpha_C}{x_A + x_B +  x_C}.
\end{equation}
This means that if the slope of $A\cup B$, $\alpha_{A\cup B}$, is higher or equal  then the slope of $A\cup B\cup C$, $\alpha_{A\cup B\cup C}$, then a continuous increment of $\alpha_A$ will lead to a continuous maximization of $S(A\cup B\cup C)- S(A\cup B)$, until $\alpha_A$ reaches its limiting value of 1. Thus we always end up with $\alpha_A=1$.

This rule that higher slope for an interval means higher rate of change for its proper length when  compared with a second interval sharing the same end point, under a variation of the associated $\alpha$-angle, is general and applies even when both slopes are negative, just considering absolute values. If  $\alpha_{A\cup B}< \alpha_{A\cup B\cup C}$  the effect of varying $\alpha_A$ would be undetermined, unless we knew the exact form of $S(l)$.

After having sent $\alpha_A\rightarrow 1$, we focus on $\alpha_C$ in order to maximize $S(A\cup B\cup C)- S(B\cup C)$. We can have either positive, null or negative $\alpha_{B\cup C}$; if $\alpha_{B\cup C}\geq 0$ then $\alpha_{A\cup B\cup C}> 0$  and it can be either higher or lower; if $\alpha_{B\cup C}< 0$ the sign of $\alpha_{A\cup B\cup C}$ may be positive or negative ( in which case its absolute value is certainly lower than the one of $\alpha_{B\cup C}$ ).  In all the above cases but one, sending  $\alpha_C\rightarrow -1$ will result in stricter SSA2 inequality, either because of contemporary maximization of $l(A\cup B\cup C)$ and minimization of $l(B\cup C)$, or because the rate of decrease for $l(B\cup C)$ is higher then for $l(A\cup B\cup C)$; the only exception being when $\alpha_{B\cup C}> 0$ but lower then $\alpha_{A\cup B\cup C}$. 
Thus we are led to two cases
\[
\alpha_{A\cup B\cup C} > \alpha_{B\cup C} > 0  \Rightarrow \alpha_A=1,\alpha_C=undet.
\]
\[
all\;other\;cases\Rightarrow \alpha_A=1 ,\alpha_C=-1.
\]
The final step is to send $\alpha_C\rightarrow-1$ in the first case. As this is not possible by only acting on $\alpha_C$, as it may lead to a decrease of $S(A\cup B\cup C)- S(B\cup C)$,  we will contemporary change the value of $x_C$ in order to keep invariant $l(B\cup C)$ under the change of value of $\alpha_C$. That there is a solution for $x_C$ satisfying this condition is guaranteed by the fact that, changing $x_C$, $l(B\cup C)$ ranges between $l(B)$ and infinity. This operation keeps invariant everything in (\ref{ssa2}) but $l(A\cup B\cup C)$ that a simple computation shows increases. Thus for every set of parameters belonging to the first case above, there is another set of values with $\alpha_C\rightarrow-1$ and the new $x_C$, such that SSA2 with these new parameters gives a stricter bound.

\subsubsection{NPP}

The discussion is parallel to the previous section. We can have either a positive or negative value for $\alpha_{A\cup B}$ and $\alpha_{A\cup B\cup C}$; in all cases, but when $\alpha_{A\cup B\cup C}$ is positive and higher of  $\alpha_{A\cup B}$ ( also positive ), we can move $\alpha_A \rightarrow -1$ in order to create a stronger bound from SSA2.
On the other side, as we always have $\alpha_{B\cup C}\geq 0$ and either positive but lower, or negative $\alpha_{A\cup B\cup C}$, an increase in the value $\alpha_C\rightarrow 1$ will certainly lead to a stricter SSA2 inequality. Thus the two possibilities are
\[
\alpha_{A\cup B\cup C} > \alpha_{A\cup B} > 0  \Rightarrow \alpha_A=undet.,\alpha_C=1
\]
\[
all\;other\;cases\Rightarrow \alpha_A=-1 ,\alpha_C=1.
\]
Working in a parallel way to the corresponding PPN case we can make the first case above to $\alpha_A=-1,\alpha_C=1$, by adjusting the value of $x_A$.

\subsubsection{PPP}

If $\alpha_{A\cup B}\geq \alpha_{A\cup B\cup C}\geq 0$, then a continuous increment of $\alpha_A$ will lead to a continuous maximization of $S(A\cup B\cup C)- S(A\cup B)$, until $\alpha_A$ reaches its limiting value of 1. 
The increase of $S(A\cup B\cup C)- S(B\cup C)$ by varying $\alpha_C$ works similarly, only now we may have $\alpha_A=1$; thus we end up with any of the following cases:
\[
\alpha_{A\cup B} \geq \alpha_{A\cup B\cup C}, \; \alpha_{B\cup C} <\alpha_{A\cup B\cup C} \Rightarrow \alpha_A=1,\alpha_C=undet.
\]
\[
\alpha_{A\cup B} <\alpha_{A\cup B\cup C}, \; \alpha_{B\cup C} \geq \alpha_{A\cup B\cup C} \Rightarrow \alpha_A=undet. ,\alpha_C=1
\]
\[
\alpha_{A\cup B} <\alpha_{A\cup B\cup C},\;\alpha_{B\cup C} < \alpha_{A\cup B\cup C} \Rightarrow \alpha_A=undet.,\alpha_C=undet.
\]

It is now possible to reduce all the above to either a PPN or NPP configuration by sending the undetermined $\alpha$ ( $\alpha_C$, $\alpha_A$ and either of those for respectively the three cases ) to minus its value, while modifying the corresponding $x$ coordinate in order to keep the proper length of either $B\cup C$ or $A\cup B$ invariant, as already done in the previous cases. This operation keeps invariant everything in (\ref{ssa2}) but $l(A\cup B\cup C)$ that a simple computation shows increases ( this is not true for a general set of positive $\alpha_A,\alpha_B,\alpha_C$, but can be proved in the above listed configurations ). Thus for every set of values belonging to the three cases above, there is another set of values with $\alpha_{A(C)}\rightarrow-\alpha_{A(C)}$ and the new $x_{A(C)}$, such that SSA2 applied to these new parameters gives a more strict bound than the older case. Then respectively PPP passes to either PPN ( first case, $\alpha_A=1$ ) or NPP ( second case, $\alpha_C=1$) or either PPN or NPP, as we wish  ( third case ). From here on, the procedure of modifying parameters to strengthen the SSA2 inequality bound proceeds as in the previous corresponding sections.

\subsubsection{NPN}

Here $\alpha_{A\cup B\cup C}0$ may be positive or negative. In the first case  $\alpha_{A\cup B}\geq\alpha_{A\cup B\cup C}$  ( as well as $\alpha_{B\cup C}\geq\alpha_{A\cup B\cup C}$ ) so we can reduce $l(A\cup B)$ at an higher rate than $l(A\cup B\cup C)$ by sending $\alpha_A\rightarrow 0\rightarrow 1$ and NPN becomes PPN ( with now the positive $\alpha_{A\cup B\cup C}$ higher then $\alpha_{B\cup C}$, thus $\alpha_C$ remains undetermined ). In the second case $\alpha_{A\cup B}$ is either positive or negative; if it is positive again $\alpha_A\rightarrow 0\rightarrow 1$ ( decreasing $l(A\cup B)$ and increasing $l(A\cup B\cup C)$ until it becomes eventually flat, and then decreasing it as well but at a lower rate ) so again PPN with undetermined $\alpha_C$. If instead $\alpha_{A\cup B}<0$, its absolute value is lower then $\alpha_{A\cup B\cup C}$, then $\alpha_A$ cannot be changed and we consider  $\alpha_{B\cup C}$. If it is positive $\alpha_C\rightarrow 0\rightarrow 1$ and we obtain NPP; if it is negative we have $\alpha_{A\cup B}$ and $\alpha_{B\cup C}$ negative and their moduli lower than $\alpha_{A\cup B\cup C}$, so $\alpha_A,\alpha_C$ are undetermined:
\[
\alpha_{A\cup B\cup C} \geq 0  \Rightarrow \alpha_A=1,\alpha_C=undet. \;\;\; (PPN)
\]
\[
\alpha_{A\cup B\cup C} < 0, \;\alpha_{A\cup B}\geq 0  \Rightarrow\alpha_A=1,\alpha_C=undet.\;\;\; (PPN)
\]
\[
\alpha_{A\cup B\cup C} < 0, \;\alpha_{A\cup B}< 0, \;\alpha_{B\cup C}\geq 0  \Rightarrow\alpha_A=undet,\alpha_C=1 \;\;\;(NPP)
\]
\[
\alpha_{A\cup B\cup C} < 0, \;\alpha_{A\cup B}< 0,\;\alpha_{B\cup C}< 0  \Rightarrow\alpha_A=undet,\alpha_C=undet. 
\]
In the last case we can apply the same procedure used in the previous section, that is sending ( for example ) $\alpha_C\rightarrow-\alpha_C$ while changing $x_C$ in order to keep fixed  $l(B\cup C)$. Again it can be checked that this increases $l(A\cup B\cup C)$, and thus we end up with NPP. From here on we work as in the first two sections.

We can finally say that making SSA2 inequality stricter leads to either PPN with $\alpha_A=1,\alpha_C=-1$ or NPP with $\alpha_A=-1,\alpha_C=1$, that is the A and C segments become light-like and on the same "time side" with respect to B. The slope of B remains general and positive ( as the negative values are covered by the parity transformation (\ref{ppp}) )
\footnote{ We may try to modify $\alpha_B$ to make the SSA2 bound stronger, but the result always leads to an undetermined change for the inequality, and even the trick of modifying some of the $x$ coordinate as done before does not help ( as either there are non positive solutions or the new set of parameter leads to a weaker bound ). The analysis deals with all the single cases and it is quite boring, so we avoid presenting it in detail.}.

\subsection{SSA1}

Here the work is easier as the $B$ interval only appears on one side of the inequality. Moreover the sign on $\alpha_A$ and $\alpha_C$ only affect one side as well. This leads to the choice of PPP over any other configuration ( $l(A)$ and $l(C)$ are unaffected by $\alpha_{A(C)}\rightarrow -\alpha_{A(C)}$ but sign changes leading to the PPP configuration always decrease $l(A\cup B)$ and $l(B\cup C)$, everything else being fixed ), and $\alpha_B=1$ ( it reduces both $l(A\cup B)$ and $l(B\cup C)$ while keeping invariant $l(A)$ and $l(C)$ ). A further attempt to shift the value of  $\alpha_A,\alpha_C$ towards zero in order to decrease $S(A\cup B)-S(A)$ and $S(B\cup C)-S(C)$ fails, as the slopes of $A$ and $C$ are always lower ( $B$ is light-like ) then respectively the slopes of $A\cup B$ and $B\cup C$ ( and we do not fit the criteria of equation (\ref{der2}), nor does it work the trick of changing the value of $x_B$ to keep $l(A\cup B)$ fixed while increasing $l(A)$, as the solution for $x_B$ is in general not positive ). So the most strict bound on $S(l)$ from SSA1, with generic $x_A,x_B,x_C$, comes from the configuration $\alpha_A\geq 0,\alpha_B=1,\alpha_C\geq 0$.

\section{Vaidya computation}\label{B}

The main reference  is \cite{Callan:2012ip}. I will often refer to formulas written there while trying to remain consistent here. The metric of Vaidya is
\begin{equation}
ds^2=-(r^2 -m(v))dv^2 +2 dr dv + r^2 dx^2
\end{equation}
which is a solution with negative cosmological constant and an energy momentum tensor with non-zero component
\begin{equation}
T_{vv}=\frac{1}{2r}\partial_v m(v).
\end{equation}
We will pick $m(v)$ to be a step function centerd at $v=0$ and $T_{vv}$ a delta function. If the delta function is positive ( $m(v<0)=0,\;m(v>0)=m$ ) the metric respects the NCC ( see for example section 3.3 of \cite{Callan:2012ip} ) and it is given by AdS inside for $v<0$ and BTZ outside for $v>0$; if instead the delta is negative  ( $m(v<0)=m,\;m(v>0)=0$ ) the NCC is violated and we have BTZ inside for $v<0$ and AdS outside for $v>0$. Using a scaling symmetry  ( see eq. 3.18 of \cite{Callan:2012ip}  ) we can fix $m=1$.

The physics is better understood in $t,r$ coordinates, where $t$ is related to $v$ by the definition $v=t+\tilde{r}(r)$ with $\partial_r\tilde{r}(r)=1/f(r)$, having set $f(r)=r^2$ for AdS and $f(r)=r^2-1$ for BTZ . The solutions are
\begin{equation}\label{v}
v=t-1/r \;\; for\; AdS, \;\;v=t-Arcth(1/r)\;\; for\; BTZ \;r>1,  \;\;v=t-arcth(r)\;\; for\; BTZ \;r<1.
\end{equation}

The shell at $v=0$ moves towards the center at $r=0$ as the time $t$ increases, transforming, as it moves, AdS into BTZ if NCC is respected, or the opposite when NCC is violated. Note that the coordinates $r$ and $v$ are continuous across the shell, while $t$ has a finite discontinuity. The metric in $t,r$ coordinates is
\begin{subequations}
\begin{align}
ds^2 &= -r^2 dt^2 +\frac{dr^2}{r^2} + r^2 dx^2 \;\;\; AdS  \\
ds^2 &= -(r^2-1) dt^2 +\frac{dr^2}{r^2-1} + r^2 dx^2 \;\;\; BTZ. 
\end{align}
\end{subequations}

\subsection{NCC violated}
We want to solve the equation for a geodesic when NCC is violated ( see also Appendix A of \cite{Callan:2012ip}, while the easier case for NCC respected is treated in section 3.2.2 ). The case we will consider is when the geodesic's end points on the bulk's boundary are at the same time $t_b$, and separated by a space distance $\Delta x_B$. 

As the metric is $x$ independent there is a conserved charge $p_x$; further the metric is also $t$ independent but when crossing the shell, thus we can use also a charge E, conserved independently in AdS and BTZ. They can be computed to be
\begin{equation}\label{var}
p_x=g_{xx}\dot{x}=r^2\dot{x},\;\;\; E_{AdS}=g_{tt}\dot{t} =-r^2\dot{t},  \;\;\; E_{BTZ}=g_{tt}\dot{t} =-(r^2-1)\dot{t}  
\end{equation}
where the overdot represents a derivative with respect to the affine parameter $\tau$ defined on the space-like geodesic as
\begin{equation}
g_{\mu\nu}\partial_{\tau}x^{\mu}\partial_{\tau}x^{\nu}=g_{\tau\tau}=1.
\end{equation}
As the Lagrangian along the "time" $\tau$ is $L=\sqrt{g_{\mu\nu}\partial_{\tau}x^{\mu}\partial_{\tau}x^{\nu}}=1$ we can write the equation
\begin{equation}
0=H(\tau)=-L+p_i \dot{q}^i=-f-f^2(\dot{t})^2+r^2(\dot{x})^2+(\dot{r})^2
\end{equation}
that leads to the two equations ( the second obtained dividing by $(\dot{x})^2$ and rewriting $\left(\frac{\dot{r}}{\dot{x}}\right)^2=(\partial_x r)^2=(r')^2$ )
\begin{equation}\label{eqn1}
(\dot{r})^2=-\frac{f}{r^2}p_x^2+E^2+f\;\;\;(r')^2=-\frac{r^4 f}{p_x^2}+\frac{r^4 E^2}{p_x^2}-r^2f.
\end{equation}
There are two additional equations we need to derive relating the internal and external $r$ derivative on the shell; they are obtained by locally extremizing the geodesic by varying the intersection point with the shell, $r_c,x_c$ ( $v_c=0$ ) given two generic nearby geodesic points, one in AdS and one in BTZ \cite{Balasubramanian:2011ur}. The geodesic will experience a refraction with the difference between the two $r'$ as
\begin{equation}\label{eqn2}
r'_A-r'_B=\frac{1}{2}v' \;\;\; v'_A=v'_B=v'.
\end{equation}
The space-like geodesics with two end points on the boundary and crossing the shell ( twice ) are the ones we are interested in; these geodesics are symmetric with respect to the axis that passes through their vertex and the middle between the two end points ( as we will fix the two end points to be at the same time $t$ ). Thus also the two crossing points on the shell are at the same time $t$ ( different by the one at the boundary ), so $\dot{t}=0$ inside the shell and for BTZ $E_{B}=0$. This simplifies the computation. Equating
\begin{equation}\label{var2}
v'_A=(t'_A+r'_A\partial_r v)_{r=r_c}=\frac{\dot{t_A}}{\dot{x_A}}+r_A'/r_c^2=\frac{E_A}{p_x}+r_A'/r_c^2=v'_B=r_B'/(r_c^2-1)
\end{equation}
we have derived an equation for $E_A$:
\begin{equation}\label{ep}
E_A=p_x\left(\frac{r'_B}{r_c^2-1}-\frac{r'_A}{r_c^2}\right).
\end{equation}
Using the first of (\ref{eqn2}), with $v'=v'_B$
\begin{equation}\label{a}
r'_A=r'_B\left(1+\frac{1}{2(r_c^2-1)}\right).
\end{equation}
Plugging (\ref{a}) inside (\ref{ep}) and using the second of (\ref{eqn1}) for BTZ to find an expression for $r'_B$ depending only on $r_c,p_x$, we arrive at \footnote{The sign plus solution is selected as in all the three parameter ranges we will consider it is the only one allowed, see appendix A of  \cite{Callan:2012ip} for cases 1,2 and 3. }
\begin{equation}\label{en}
E_A=\frac{1}{2 r_c}\sqrt{\frac{r_c^2-p_x^2}{r_c^2-1}}.
\end{equation}

\begin{figure}[h]
\centering
\vspace{-0pt}
\includegraphics[width=0.6\textwidth]{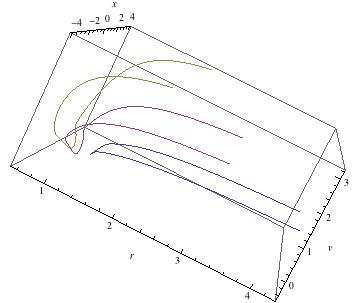}
\vspace{-0pt}
\caption{Three geodesics with parameters $(r_{c1}, r_{c2}, r_{c3}, p_{x1}, p_{x2}, p_{x3}) = (1.2, 0.75, 0.6, 1.1, 0.9, 0.8)$. Above the shell represented by the plane $v=0$ is AdS ( positive $v$), below is BTZ ( negative $v$ ). The boundary of Vaidya is at $r=\infty$, its center at $r=0$. At the shell we see the refraction, where the geodesics bend at an angle; the solution at negative $v$ is the one for BTZ, it depends on $p_x$ and a parameter $k$ that is adjusted so that this solution meets the one in AdS at the radius $r_c$ for the same two values of the affine parameter, $\tau_{c1}$ and $\tau_{c2}$. More properties will be discussed later.}
\label{figura1}
\end{figure}

 A geodesic with nonzero energy will depend on two parameters, the radius of the two crossing points with the shell $r_c$ and the momenta $p_x$, that we will soon see to coincide with the radius of the vertex of the geodesic. If the geodesic doesn't cross the shell, or if we are concentrating on the arc after the crossing, for symmetry $E=0$. That geodesic will then depend only on $p_x$ in the first case, or on $p_x$ and an additional parameter $k$ in the second. This $k$ will be then tuned to be a function of $p_x$ and $r_c$ as we want this geodesic arc, going from the shell inward, to be glued correctly to two symmetric pieces of a second geodesic coming from the boundary to the shell and having $r_c$ and the same $p_x$ as parameters. An example of three different geodesics in different parameters range is shown in figure (\ref{figura1})
 
Let us start by solving the geodesic equation for BTZ in $\dot{r}$ given by the first of (\ref{eqn1}). There are two types of solutions, either going from the boundary towards the center of BTZ and back, or starting from the center and moving towards the boundary and back, with the vertex that will be either at smaller or higher $r$ then $r_c$; further its value will be respectively $p_x$ for $p_x\geq 1$ or one for $p_x<1$ in the first case and one for $p_x\geq 1$ or $p_x$ for $p_x<1$ in the second ( also note that the first solution is (3.36) of \cite{Callan:2012ip} with $E=0$ when $c=\log p_x$. ).
\begin{subequations}
\begin{align}
r_{B,1}(\tau,p_x,k_1)&=& \frac{1}{2 p_x e^{(k_1 + \tau)}} \sqrt{ e^{4 (k_1  + \tau)} + p_x^4 (-1 + p_x^2)^2 + 2 e^{2 (k_1  + \tau)} p_x^2 (1 + p_x^2)} \label{rb1} \\
r_{B,2}(\tau,p_x,k_2)&=& \frac{1}{2} \sqrt{e^{-2 (\tau + k_2)} (-2^{4 (\tau + k_2)} - (-1 + p_x^2)^2 + 
   2 e^{2 (\tau + k_2)} (1 + p_x^2)) }. \label{rb2} 
\end{align}
\end{subequations}
The values of $\tau_{c1}$ and $\tau_{c2}$ at which the geodesic reaches the shell at $v=0$ and $r=r_c$ will be given later; The value of $\tau_0$ when the geodesic reaches its vertex is instead
\begin{equation}
\tau_0= - k_{1,2} + \frac{1}{4} \log((1 - p_x^2)^2).
\end{equation}

More complicated is the solution for the AdS part of the geodesic, as the energy (\ref{en}) is here nonzero. The solution depends on the range of the parameters $r_c,p_x$; here we list the relevant formulas
\begin{enumerate}
\item $r_c > p_x > 1$: we solve the first equation of (\ref{eqn1}) for the piece of the geodesic on the right of the BTZ solution ( the choice $"+"$ for the sign will be explained shortly ):
\begin{equation}\label{ra1}
\dot{r}_A=\sqrt{r_A^2+E_A^2-p_x^2} \;\;\;\; r_A=\frac{1}{2}e^{-\tau}\left(e^{2\tau}+p_x^2-E_A^2\right)
\end{equation}
it goes from the shell in $r_A=r_c$ at
\begin{equation}
\tau_{c2}=\log\left(r_c+\sqrt{r_c^2+E_A^2-p_x^2}\right)
\end{equation}
to the boundary at $\tau\rightarrow \infty$. As $v'|_{v=0}$ should always be positive ( because the metric is $x$-independent having a non monotonically behaviour of $v(x)$ from the vertex till the boundary, can only increase the length ) and from (\ref{var2}) $v'=r_B'/(r_c^2-1)$, we have that $r_B'>0$ as well, thus selecting the BTZ solution $r_{B,1}$. Further the relation (\ref{a}) between $r_A'>0$ and $r_B'>0$ tells us that also $r_A'>0$ for the present range of parameters; this explains also the choice of sign in (\ref{ra1}) as we conventionally consider an affine parameter $\tau$ increasing from left to right along the geodesic. The value of $k_1$ is fixed by requiring  $r_{B,1}(\tau_{c2})=r_c$:
\begin{equation}\label{k1}
k_1=\frac{1}{2} \log\left(-\frac{p_x^2  (1 + p_x^2 - 2 r_c^2 +(1 - r_c^2) 4 r_c E_A)}{(E_A - r_c (1 + 2 r_c E_A))^2}\right).
\end{equation}
The second arc in AdS, going from the boundary at $\tau\rightarrow-\infty$ to the shell at $\tau_{c1}$, can be easily obtained by symmetry. The value of $\tau_{c1}$ is simply given by the relation $\tau_{c1}+\tau_{c2}=2\tau_0$ and can be checked to produce again $r_c$ when inserted inside both $r_{B,1}$ and $r_{A}$. The values of  $x(\tau)$ and $v(\tau)$ are obtained respectively from integrating the first of (\ref{var}) ( with the integration constant conventionally fixed so that $x(\tau_0)=0$  ) and from the definition of $v$ (\ref{v}), where $t(\tau)$ again comes from integrating the second of (\ref{var}) with integration constant such that $t(\tau_{c1})=t(\tau_{c2})=1/r_c$ ( for BTZ $t=t_c=arcth(1/r_c)$ and we see the finite discontinuity of $t$ in passing through the shell ).
We can also compute the boundary time $t_b$ and distance $\Delta x_B$ of the two ending points of the geodesic, as a function of $p_x,r_c$:
\begin{equation}\label{dxb}
\Delta x_{b}=2 \left(-\frac{p_x}{2 r_c^2} AppellF1\left(1,\frac{1}{2}, \frac{1}{2}, 2, \frac{1}{r_c^2}, \frac{p_x^2}{r_c^2}\right) + 
   arcth\left(\frac{1}{p_x}\right) + \Delta x_{sb}\right)
\end{equation}
with AppellF1 a kind of Hypergeometric function coming from the integration of $x$ between the vertex and the shell and $\Delta x_{sb}$ the distance in $x$ between the crossing with the shell and the boundary point
\begin{equation}\label{dxsb}
\Delta x_{sb}=\frac{p_x}{r_c^2 + r_c \sqrt{r_c^2 + E_A^2 - p_x^2}},
\end{equation}
and
\begin{equation}\label{tb}
t_b=\frac{1}{r_c} + \frac{E_A}{p_x }\Delta x_B.
\end{equation}
We will also need to reverse the formula (\ref{tb}) to obtain the value of $p_x$ as function of $r_c$ and a given $t_b$.
\begin{equation}\label{pxt}
p_x= r_c \frac{\sqrt{4 - 4 r_c t_b + t_b^2}}{|2 r_c + t_b - 2 r_c^2 t_b|}.
\end{equation}
Before going on let me comment that this last equation (\ref{pxt}) does not work for any value of $t_b$ and $r_c$, but there is just a restricted range for which the geodesic parameterized by $r_c,p_x(t_b,r_c)$ correctly has boundary time equal to $t_b$. We will take this into account when plotting parameter regions of geodesics matching some fixed boundary time.  
\item $1> r_c > \sqrt{1/2} ,\;1>p_x > r_c $: equation and solution for $r_{A}$ are the same as the case above. Now however we need to glue the BTZ solution $r_{B,2}$ ( positive $v'$ means now $r_A'>0$ and $r_B'<0$ ) . Working as before we obtain 
\begin{equation}\label{k22}
k_2=\frac{1}{2} \log\left(\frac{1 + p_x^2 - 2 r_c^2 +(1 - r_c^2) 4 r_c E_A}{(E_A - r_c (1 + 2 r_c E_A))^2}\right).
\end{equation}
 $\tau_{c2}$ is as before and $\tau_{c1}$ can be computed using the new $k_2$ value inside $\tau_0$. $x(\tau)$ and $v(\tau)$ are computed as in the previous case as the equations for $t_b$, $\Delta x_{b}$,            $\Delta x_{sb}$ and $p_x(t_b)$ are the same as  (\ref{dxb}), (\ref{dxsb}), (\ref{tb}) and (\ref{pxt}), even in this parameter range.
\item $\sqrt{1/2}> r_c >0, \;p_x^2 - E_A^2 > 0$ that is $p_x < r_c/(1 - 2 r_c^2)$: again a sign switch with both $r_A'<0$ and $r_B'<0$. Thus the equation for $\dot{r}_A$ picks up the negative sign, and accordingly changes the solution
\begin{equation}\label{ra3}
\dot{r}_{A}=-\sqrt{r_A^2+E_A^2-p_x^2} \;\;\;\; r_A=\frac{1}{2}e^{\tau}\left(e^{-2\tau}+p_x^2-E_A^2\right),
\end{equation}
\begin{equation}
\tau_{c2}=-\log\left(r_c+\sqrt{r_c^2+E_A^2-p_x^2}\right)
\end{equation}
\begin{equation}\label{k23}
k_2=\frac{1}{2} \log\left((1 + p_x^2 - 2 r_c^2 +(1 - r_c^2) 4 r_c E_A)(E_A + r_c (1 -2 r_c E_A))^2\right).
\end{equation}
$x(\tau)$ and $v(\tau)$ are computed as in the previous cases. Now we have slightly different formulas for $\Delta x_b$, $\Delta x_{sb}$
\begin{equation}\label{dxb2}
\Delta x_{b}=2 \left(-\frac{p_x}{2 r_c^2} AppellF1\left(1,\frac{1}{2}, \frac{1}{2}, 2, \frac{1}{r_c^2}, \frac{p_x^2}{r_c^2}\right) + 
   \Theta\left(\frac{1}{p_x}-1\right) arcth\left(p_x\right)+ 
  \Theta(p_x-1) arcth\left(\frac{1}{p_x}\right) + \Delta x_{sb}\right)
\end{equation}
\begin{equation}\label{dxsb2}
\Delta x_{sb}=\frac{p_x (r_c + \sqrt{r_c^2 + E_A^2 - p_x^2})}{r_c (-E_A^2 + p_x^2)},
\end{equation}
( $\Theta$ is the usual step function ) while (\ref{tb}) and (\ref{pxt}) for $t_b$ and $p_x(t_b)$ still apply if using the new quantity $\Delta x_{sb}$ above.
\item $\sqrt{1/2}> r_c >0, \;p_x^2 - E_A^2 < 0$: not interesting as the geodesic never goes to the boundary. 
\end{enumerate}
The goal is to check weather two "concentric" geodesics linked to the end points of the intervals $B$ and $A\cup B\cup C$ respectively, all at the same fixed boundary time $t_b$ and choosing $A$ to be symmetric to $C$, develop or not a time-like distance in the interior. As the generic problem is complicated, we will limit to check the distance between the vertices ( that, given the space-like boundary conditions are the most likely to be time-like separated ). As the two vertices are at the same $x$, the problem will be in the $r-v$ plane. The first step is to consider two generic points in this plane, we will call $(r_i,v_i)$ the coordinates of the point with the lowest value of $r$  and $(r_f,v_f)$ the other, $r_i<r_f$ ( if $r_i=r_f$ and $v_i\neq v_f$ the distance is obviously time-like ). The procedure will be to shoot out of $(r_i,v_i)$ two light-like geodesics inside the plane $x=0$ and towards the boundary ( increasing r ) bordering a space-like region. Then we will check if $(r_f,v_f)$ is included inside this region or not. 

\begin{figure}[h]
\centering
\vspace{-0pt}
\includegraphics[width=1\textwidth]{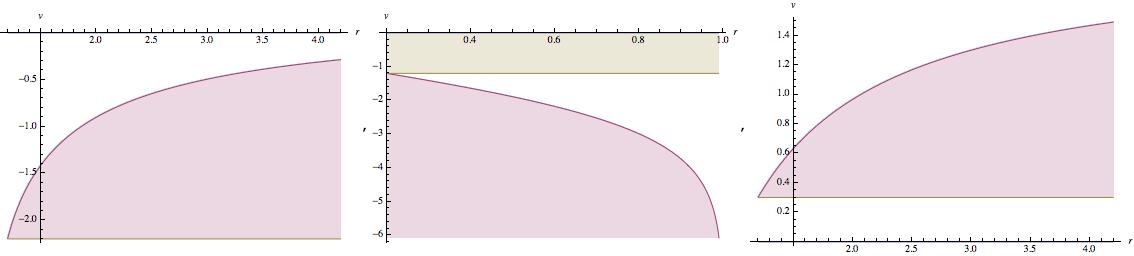}
\vspace{-10pt}
\caption{Plot of space-like regions in BTZ for $r>1$ and $r<1$ and AdS. The picture for AdS is similar to the first plot, just the curve is not given by a log but by 1/r. $(r_i,v_i)$ is at the intersection of the two light-like geodesics with the space-like region in between ( or above and below as in the second graph ) coloured.}
\label{figura2b}
\end{figure}

We will start considering both $(r_i,v_i)$ and $(r_f,v_f)$ past the shell, inside BTZ and, for the moment, both belonging to the same parameter range, among the ones listed above ( this will be generalized later ). Then the relevant metric is
\begin{equation}
ds^2 = -(r^2-1) dt^2 +\frac{dr^2}{r^2-1}.
\end{equation}
A light-like geodesic has $ds=0$, so $dt^2=dr^2/(r^2-1)^2$ and the space-like region is either $dt^2<dr^2/(r^2-1)^2$ if $r>1$ or $dt^2>dr^2/(r^2-1)^2$ if $r<1$. In the $v,r$ coordinates we are using, with $dv=dt+dr/(r^2-1)$ this means, as $dr>0$ by assumption, the line element is space-like iff
\begin{subequations}
\begin{align}
dv>0  \;\;\;and\;\;\; dv &<2\frac{dr}{r^2-1} \;\;\;\; r>1\label{btzll1} \\
dv>0  \;\;\;or\;\;\; dv &<2\frac{dr}{r^2-1} \;\;\;\; r<1.\label{btzll2}
\end{align}
\end{subequations}
If instead both  $(r_i,v_i)$ and $(r_f,v_f)$ belong to AdS \footnote{Soon we will consider the mixed case.}
\begin{equation}\label{adsll}
0<dv<2\frac{dr}{r^2}.
\end{equation}
Integrating between $r_i$ and $r_f$ we have the space-like-type-inequalities for BTZ ( the log can also be expressed as an arcth, as one wishes ) singling out either one connected region for $r_f,r_i>1$ or two disconnected ones for $r_f,r_i<1$:
\begin{subequations}
\begin{align}
v_f-v_i>0 \;\;\;and\;\;\; v_f-v_i &<\log[\frac{(-1 + r_f) (1 + r_i)}{(1 + r_f) (-1 + r_i)}]  \;\;\;\; r_i,r_f>1\label{btzc1} \\
v_f-v_i>0 \;\;\;or\;\;\; v_f-v_i &<\log[\frac{(-1 + r_f) (1 + r_i)}{(1 + r_f) (-1 + r_i)}]  \;\;\;\; r_i,r_f<1.\label{btzc2} 
\end{align}
\end{subequations}
Obviously one has to check that with the parameters values considered, not only the two vertices both belong to BTZ, but also the light-like geodesic doesn't cross the shell. The crossing case will be shown soon. For AdS instead there is always a single connected space-like region
\begin{equation}\label{adsc}
0<v_f-v_i<-\frac{2}{r_f} + \frac{2}{r_i}.
\end{equation}
The three cases are shown in figure \ref{figura2b}.
Two cases are missing, that is when $r_i<1$ and $r_f>1$, thus crossing the BTZ horizon and having two different parameter-range geodesics, and when one vertex is in BTZ and the other in AdS. To determine the light-like geodesics the recipe is simply to integrate the null line element for the correct domain. So if $r_i<1$ and $r_f>1$ we need to integrate $0$ and $2 dr/(r^2-1)$ until $r=1$ and then from there to $r_f$; the integral around $r=1$ is divergent but the sum can be easily regularized ( integrating from $r_i$ until $1-\epsilon$ and from $1+\epsilon$ to $r_f$ and then sending $\epsilon\rightarrow 0$ ) obtaining the light-like distance between $v_i$ and what we may call $v_l$ ( the $v$ coordinate of the light-like geodesic at $r=r_f$ ), to be:
\begin{equation}\label{mxd1c}
v_l=v_i \;\;\; or \;\;\; v_l=v_i+\log[\frac{(-1 + r_f) (1 + r_i)}{(1 + r_f) (-1 + r_i)}].
\end{equation}
When instead we have a light-like geodesic crossing the shell, we have first to obtain the crossing point $r_{cl}$, and then integrate accordingly in the two different regions ( before and past $r_{cl}$ ). The formula for $r_{cl}$ is ($v_{cl}=0$)
\begin{equation}
 -v_i=\int_{r_i}^{r_{cl}}\frac{2dr}{r^2-1}=\log[\frac{(-1 + r_{cl}) (1 + r_i)}{(1 + r_{cl}) (-1 + r_i)}]
\end{equation}
so
\begin{equation}\label{mxd2c2}
r_{cl}=\frac{-1 + e^{-v_i} - r_i - e^{-v_i} r_i}{-1 - e^{-v_i} - r_i + e^{-v_i} r_i}
\end{equation}
and ( here necessary $r>1$ otherwise this geodesic goes towards smaller $v$ and never crosses the shell starting from BTZ; also  the complementary case is not possible when, starting from a point in AdS we have a light-like geodesic that crosses the shell, as they always goes for higher values of $v$. )
\begin{equation}\label{mxd2c}
0<v_f-v_i<-\frac{2}{r_f} + \frac{2}{r_{cl}}.
\end{equation}
\begin{figure}[h]
\centering
\vspace{-0pt}
\includegraphics[width=0.8\textwidth]{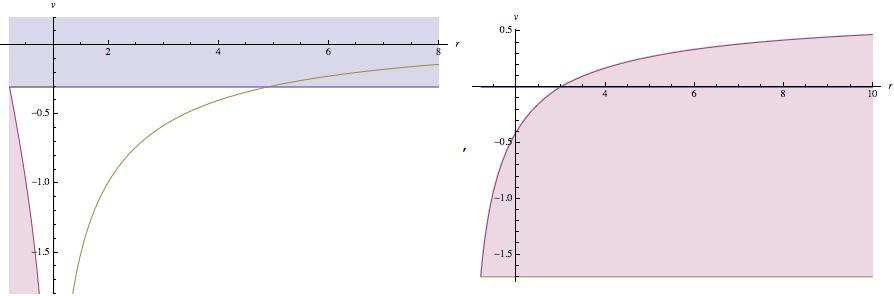}
\vspace{-0pt}
\caption{Plot of two space-like regions. In the first we start at $r_i=0.2,t_i=-0.1$ and move in the $r>1$ region inside BTZ; the lower space-like region when $r<1$ does not extend to $r>1$, as there is not such a space-like geodesic crossing the horizon. In the second we focus on the crossing of the shell starting from $r_i=1.2,t_i=-0.5$ that happens at at $r_{cl}=2.977$.}
\label{figura4}
\end{figure}
Let us now apply these inequalities to study in which parameter range the distance between the two geodesic vertices becomes time-like. As the boundary time has to be the same for both geodesics, the four parameters describing them are not independent. In particular we can freely choose a couple $r_{c1},p_{x1}$ plus $r_{c2}$ and then use equation (\ref{pxt}) to determine $p_{x2}(t_b(r_{c1},p_{x1}),r_{c2})$, so that $t_{b1}=t_{b2}$. 
For simplicity we will restrict our analysis to situations were the two geodesics belong to the same case, even if a more general computation is possible following the above discussion.

\subsubsection{Case 2: $1> r_c > \sqrt{1/2} ,\;1>p_x > r_c $}
We can easily evaluate $v_f-v_i$, first by calling "f" the geodesic with higher $p_x$, let us fix it to be $p_{x1}$ ( this will become a condition on the three parameters $r_{c1},p_{x1},r_{c2}$, if not respected just switch names ) and "i" the other. Then, as the vertex $r$-value of a geodesic is always $p_x$, with $v(r_c)=0$ by definition:
\begin{equation}
v_f-v_i=v_1-v_2=-\int_{p_{x1}}^{r_{c1}}dr \partial_{r}v+\int_{p_{x2}}^{r_{c2}}dr \partial_{r}v=\frac{1}{2} \log\left(\frac{(-1 + p_{x1}) (1 + r_{c1})(1 + p_{x2}) (-1 +
       r_{c2})}{(1 + p_{x1}) (-1 + r_{c1})(-1 + p_{x2}) (1 + r_{c2})}\right).
\end{equation}
So we end up with (\ref{btzc2}) to become
 \begin{subequations}
\begin{align}
\frac{1}{2}\log\left(\frac{(-1 + p_{x1}) (1 + r_{c1})(1 + p_{x2}) (-1 + r_{c2})}{(1 + p_{x1}) (-1 + r_{c1})(-1 + p_{x2}) (1 + r_{c2})}\right) &>0  \;\;\;\; or \\
\frac{1}{2}\log\left(\frac{(-1 + p_{x1}) (1 + r_{c1})(1 + p_{x2}) (-1 + r_{c2})}{(1 + p_{x1}) (-1 + r_{c1})(-1 + p_{x2}) (1 + r_{c2})}\right) &<\log\left(\frac{(-1 + p_{x1}) (1 + p_{x2}) }{(1 + p_{x1}) (-1 + p_{x2})}\right) .
\end{align}
\end{subequations}
As the log is monotonically increasing we can obtain the parameters region for space-like distance between the vertices to be given by either one or the other of the following inequalities
 \begin{subequations}
\begin{align}
y(p_{x1}) y(r_{c2}) &< y(p_{x2}) y(r_{c1})   \label{in21}\\
& or \nonumber \\
y(r_{c1}) y(p_{x1}) &< y(p_{x2}) y(r_{c2})\label{in22} 
\end{align}
\end{subequations}
where we have defined
\begin{equation}
y(r)=\frac{1 + r}{-1 + r}.
\end{equation}
The convention we made that $p_{x1}>p_{x2}(r_{c1},p_{x1},r_{c2})$ can be shown to be equivalent to $r_{c1}>r_{c2}$, and we also should impose $p_{x1}>r_{c1}$. Further the parameter space $1>p_{x1}>r_{c1}>r_{c2}> \sqrt{1/2}$ should be additionally constrained by $t_b(r_{c1},p_{x1})-t_b(r_{c2},p_{x2})=0$, that we have seen is fulfilled by the definition of 
$p_{x2}$ only on a restricted parameter range; this will be implemented numerically in all the examples. We can sweep this parameter space to look for what volume satisfies the inequalities (\ref{in21}) and  (\ref{in22}) and what violates them. The result is that they are always violated as shown in figure \ref{figura5}

\subsubsection{Case 3: $\sqrt{1/2}> r_c >0, \;p_x^2 - E_A^2 > 0$}
This works more or less as the previous section, just the parameter space is modified. The inequalities for the distance between the vertices $1,2$ to be space-like are as previously:
 \begin{subequations}
\begin{align}
y(p_{x1}) y(r_{c2}) &< y(p_{x2}) y(r_{c1})  \label{in31}\\
& or \nonumber \\
y(r_{c1}) y(p_{x1}) &< y(p_{x2}) y(r_{c2}).\label{in32} 
\end{align}
\end{subequations}
Again $p_{x1}>p_{x2}(r_{c1},p_{x1},r_{c2})$ is equivalent to $r_{c1}>r_{c2}$, and thus the parameter space is $p_{x1}>r_{c1}>r_{c2}$ plus the additional constrain $p_{x1}^2 - E_{A1}^2 > 0$ that is $p_{x1} < r_{c1}/(1 - 2 r_{c1}^2)$ that further restricts it, plus the usual $t_b(r_{c1},p_{x1})-t_b(r_{c2},p_{x2})=0$. Again always violation, as shown in figure \ref{figura5b} 

\subsubsection{Case 1: $r_c > p_x > 1$}
This is the most interesting situation as we will end up with both space-like and time-like vertices. Here the choice $p_{x1}>p_{x2}(r_{c1},p_{x1},r_{c2})$ does not translate to an easy condition on the parameter space, so it is more convenient to avoid it and independently deal with the cases $p_{x1}>p_{x2}$ and $p_{x1}<p_{x2}$. The inequalities representing space-like distances between the vertices thus are:
\begin{subequations}
\begin{align}
y(r_{c1}) y(p_{x1}) &> y(p_{x2}) y(r_{c2})  \;\;\;and \;\;\; y(p_{x1}) y(r_{c2}) > y(p_{x2}) y(r_{c1})  \;\;\;p_{x1}<p_{x2}\label{in11}\\
y(r_{c2}) y(p_{x2}) &> y(p_{x1}) y(r_{c1}) \;\;\;and \;\;\; y(p_{x2}) y(r_{c1}) > y(p_{x1}) y(r_{c2})  \;\;\;p_{x1}>p_{x2}.\label{in12}
\end{align}
\end{subequations}
The parameter range for $r_{c1},p_{x1},r_{c2}$ is then limited by the three constraints $r_{c1}>p_{x1}$, $r_{c2}>p_{x2}(r_{c1},p_{x1},r_{c2})>1$ and  $t_b(r_{c1},p_{x1})-t_b(r_{c2},p_{x2})=0$. Inside this volume we separately deal with the subspaces $p_{x1}>p_{x2}$ and $p_{x1}<p_{x2}$; now we can apply (\ref{in11})  and (\ref{in12}) on the relevant space and see if they are satisfied or not. It turns out that when  $p_{x1}<p_{x2}$, $y(p_{x1}) y(r_{c2}) > y(p_{x2}) y(r_{c1}) $ is always respected, while $y(r_{c1}) y(p_{x1}) > y(p_{x2}) y(r_{c2})$ is respected in some region and violated in its complementary, as is shown in figure \ref{figura7}.

On the other side when $p_{x1}>p_{x2}$ $y(r_{c2}) y(p_{x2}) > y(p_{x1}) y(r_{c1})$ is always respected, while $y(p_{x2}) y(r_{c1}) > y(p_{x1}) y(r_{c2}) $ is respected in some region and violated in its complementary ( with respect to the total space  ). This is shown in figure \ref{figura8}.

\subsection{NCC respected}
Now we have BTZ outside and AdS inside the shell. Skipping the derivation, that is easier then the one in the previous sections and  can be found in \cite{Callan:2012ip}, we limit ourselves in listing the relevant formulas for our purposes. The boundary time is 
\begin{equation}\label{tbncc}
t_b=\frac{1}{2} \log\frac{(A_{-}+ e^{2\tau_{c2}}) (r_c + 1)}{(A_{+}+ e^{2\tau_{c2}}) (r_c -  1)}
\end{equation}
with
\[
A_{\pm}\equiv p_x^2-(1\pm E_{B})^2 \;\;\;\;\;\;E_{B}=-\frac{1}{2 r_c^2}\sqrt{r_c^2-p_x^2}
\]
\[
\tau_{c2(1)}=\frac{1}{2}\log\left(-\frac{1}{2}(A_+ +A_-)-2+4r_c^2\pm\sqrt{-4A_+ A_- +(A_+ +A_- +4-4r_c^2)^2}\right).
\]
The equivalent of (\ref{pxt}) for the above $t_b$ is instead given by 
\begin{equation}\label{pxads}
p_x=\frac{r_c \sqrt{-\tanh(t_b)} \sqrt{ 4 r_c - (1 + 4 r_c^2) \tanh(t_b)}}{|-2 r_c +  \tanh(t_b) + 2 r_c^2 \tanh(t_b)|}.
\end{equation}
Finally the total $x$ coordinate length covered by the geodesic at the boundary can be computed to be\footnote{There is a typo of an additional $\frac{1}{2}$ in the corresponding formula for $l_x$ of \cite{Callan:2012ip} in front of the log.}
\begin{equation}\label{xbncc}
\Delta x_b=\frac{2}{r_c p_x}\sqrt{r_c^2-p_x^2}+\log\left(\frac{B_++e^{2\tau_{c2}}}{B_-+e^{2\tau_{c2}}}\right)\;\;\;\;\;\;B_{\pm}\equiv (p_x\pm 1)^2-E_{B}^2.
\end{equation} 
With this we can repeat the analysis of the previous sections; for geodesics crossing the shell and whose vertices are in AdS, the difference between their $v$-coordinates becomes ( $p_{x1}<p_{x2}$ and $t=const$ in AdS for the usual symmetry argument )
\begin{equation}
v_2-v_1=-\int_{p_{x2}}^{r_{c2}}dr \partial_{r}v+\int_{p_{x1}}^{r_{c1}}dr \partial_{r}v=-\frac{1}{p_{x2}} + \frac{1}{p_{x1}} + \frac{1}{r_{c2}} - \frac{1}{r_{c1}}
\end{equation}
and the space-like condition is 
\begin{equation}
-\frac{2}{p_{x2}} + \frac{2}{p_{x1}}>-\frac{1}{p_{x2}} + \frac{1}{p_{x1}} + \frac{1}{r_{c2}} - \frac{1}{r_{c1}}>0
\end{equation}
or more easily
\begin{equation}\label{csl}
\frac{1}{p_{x1}} + \frac{1}{r_{c1}}>\frac{1}{p_{x2}} + \frac{1}{r_{c2}}  \;\;\;\; and \;\;\;\; \frac{1}{p_{x1}} + \frac{1}{r_{c2}}>\frac{1}{p_{x2}} + \frac{1}{r_{c1}}.
\end{equation}
Given $p_{x2}=p_{x2}(t_b(r_{c1},p_{x1}),r_{c2})$ by (\ref{pxads}) ( restricting to the parameter space such that $p_{x1}<p_{x2}$, and with the corresponding version of (\ref{csl}) given by $1\leftrightarrow 2$ on the complementary $p_{x1}>p_{x2}$ ), we can already visualize the parameter volume that respect the first and second of (\ref{csl}). Again we also implement the constrain on the parameter space such that the difference between $t_b(r_{c1},p_{x1})$ and $t_b(r_{c2},p_{x2}(r_{c1},p_{x1},r_{c2}))$ is smaller then some arbitrarily small number. Numerical results in random parameter domains show that vertices are always at space-like distance one from the other.

\end{document}